\documentclass[11pt,letterpaper]{article}
\pdfoutput=1

\usepackage{graphicx,array}
\usepackage[dvipsnames]{xcolor}
\definecolor{darkblue}{rgb}{0,0.1,0.5}
\definecolor{darkgreen}{rgb}{0,0.5,0.2}
\definecolor{seablue}{rgb}{0,0.2,0.6}
\usepackage{latexsym}
\usepackage{amssymb, amsmath}
\usepackage{bm}        
\usepackage[numbers,sort&compress]{natbib}
\usepackage{bm,relsize}
\usepackage{slashed}
\usepackage{mathrsfs}
\usepackage{hyperref} 
\hypersetup{
    colorlinks=true,       
    linkcolor=darkblue,          
    citecolor=BrickRed,        
    filecolor=magenta,      
    urlcolor=MidnightBlue           
}
\usepackage[all]{hypcap} 
\usepackage{subcaption}

\usepackage[font=small,labelfont=bf,labelsep=period]{caption}

\newcommand{\GeV}{\mathrm{GeV}}
\newcommand{\TeV}{\mathrm{TeV}}

\newcommand{\be}{\begin{equation}}
\newcommand{\ee}{\end{equation}}

\newcommand{\fb}{\mathrm{fb}}

\begin{document}

\begin{flushright}
DESY 18-116
\end{flushright}
\vspace{.6cm}
\begin{center}
{\LARGE \bf Fusing Vectors into Scalars\\  \vspace{3mm} at High Energy Lepton Colliders}
\bigskip\vspace{1cm}{

Dario Buttazzo$^a$,\, Diego Redigolo$^{b,c,d}$,\, Filippo Sala$^{e}$ and Andrea Tesi$^f$ }
\\[7mm]
 {\it \small
$^a$INFN Sezione di Pisa, Largo B. Pontecorvo 3, I-56127 Pisa, Italy\\
$^b$School of Natural Sciences, Institute for Advanced Study, Einstein Drive, Princeton, NJ 08540, USA, \\ 
$^c$Raymond and Beverly Sackler School of Physics and Astronomy, Tel-Aviv University, Tel-Aviv 69978, Israel, \\
$^d$Department of Particle Physics and Astrophysics, Weizmann Institute of Science, Rehovot 7610001,Israel \\ 
$^e$ DESY, Notkestra$\beta$e 85, D-22607 Hamburg, Germany\\
$^f$INFN Sezione di Firenze, Via G. Sansone 1, I-50019 Sesto Fiorentino, Italy\\
 }

\end{center}

\bigskip \bigskip \bigskip \bigskip

\centerline{\bf Abstract} 
\begin{quote}
We study vector boson fusion production of new scalar singlets at high energy lepton colliders. We find that CLIC has the potential to test single production cross-sections of a few tens of attobarns in di-Higgs and di-boson final states. In models with a sizeable singlet-Higgs mixing, these values correspond to a precision in Higgs couplings of order 0.1$\%$ or better. We compare our sensitivities with those of the LHC and interpret our results in well-motivated models like the Twin Higgs, the NMSSM and axion-like particles. Looking forward to even higher energy machines, we show that the reach of muon colliders like LEMMA or MAP overcomes the one of future hadron machines like FCC-hh. We finally study the pair production of the new scalar singlets via an off-shell Higgs. This process does not vanish for small mixings and will constitute a
crucial probe of models generating a first order electro-weak phase transition. 
\end{quote}

\vfill
\noindent\line(1,0){188}
{\scriptsize{ \\ E-mail:\texttt{ \href{mailto:dario.buttazzo@pi.infn.it}{dario.buttazzo@pi.infn.it}, \href{d.redigolo@gmail.com}{d.redigolo@gmail.com}, \href{mailto:filippo.sala@desy.de}{filippo.sala@desy.de}, \href{andrea.tesi@fi.infn.it}{andrea.tesi@fi.infn.it}}}}
\newpage

\tableofcontents

\section{Introduction}

The legacy of the LHC on the Standard Model (SM) Higgs boson will not be enough to fully assess the structure of the Higgs sector.
This provides a strong motivation for future lepton colliders, which is somehow independent of other possible LHC findings.
An additional important motivation for these machines comes from their potential to directly produce and test new physics states.
It is the purpose of this paper to study the interplay of these two motivations in a concrete yet general model, focussing on high-energy lepton colliders (HELCs).

\begin{figure}[t]
\centering
\includegraphics[width=.7\textwidth]{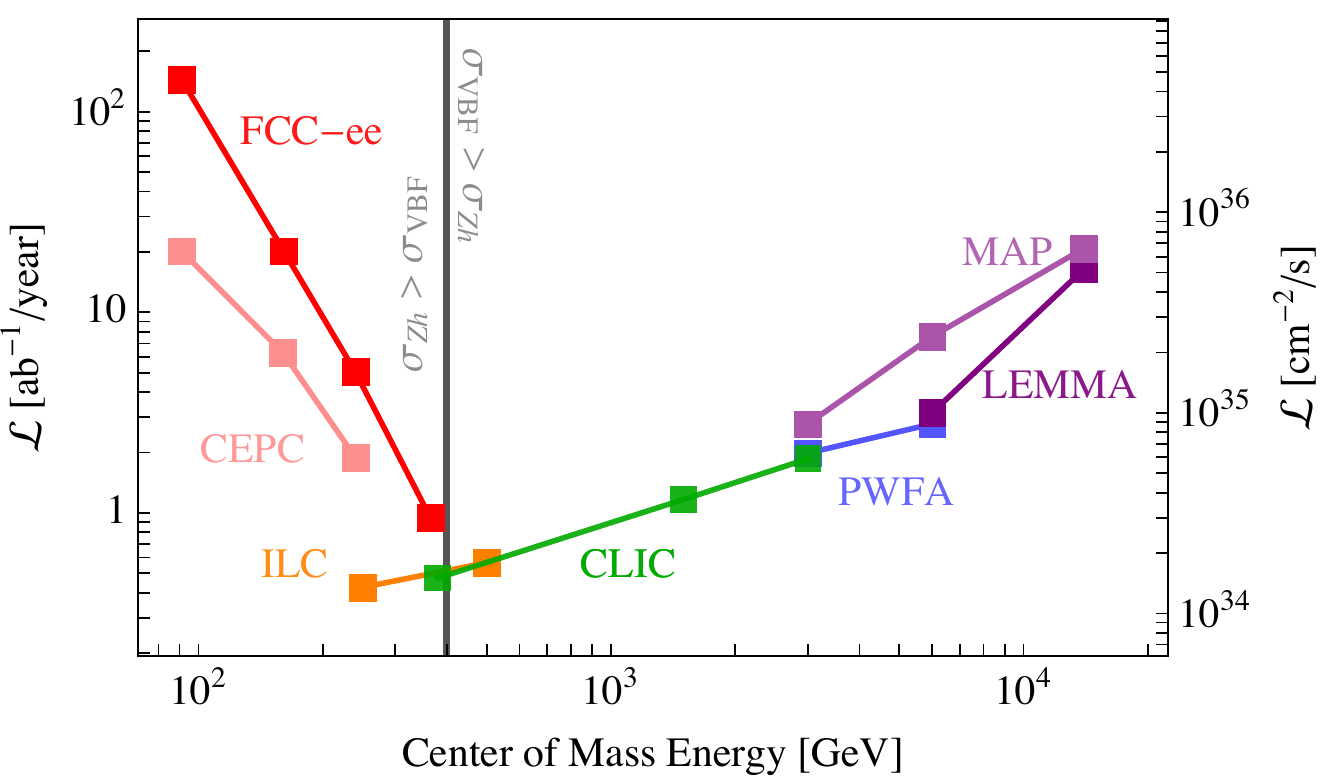} 
\caption{Various proposals for future lepton colliders as a function of center-of-mass energy and instantaneous luminosity $\mathcal{L}$, where we have assumed one interaction point for the linear ones (ILC, CLIC, PWFA) and two for the circular ones (FCC-ee, CEPC, LEMMA, MAP).
The total integrated luminosity can be obtained multiplying the left axis by the years of run and by the fraction of a year in which a machine will be running.
The luminosities reported correspond to luminosities at the peak of energy only for circular colliders.
On the left/right of the vertical grey line the Higgs-strahlung rate is larger/smaller than the vector boson fusion one.}
\label{fig:lep_colliders}
\end{figure}

The different proposals of lepton colliders can be classified in low-energy, such as FCC-ee~\cite{Gomez-Ceballos:2013zzn,fcc-ee-lumi}, CEPC~\cite{CEPC-SPPCStudyGroup:2015csa,cepc-lumi}, and ILC in its current design~\cite{Baer:2013cma,Evans:2017rvt}, and high-energy, such as CLIC~\cite{Linssen:2012hp} in its stages at a center-of-mass energy of 1.5 TeV (Stage II) and 3 TeV (Stage III).
Other futuristic examples of HELCs are high-energy circular muon colliders to be possibly built at CERN~\cite{Shiltsev:2018qbd} and/or at the muon accelerator facility at Fermilab~\cite{Delahaye:2013jla}.
Muons could be produced from $p$ scattering on a target (MAP~\cite{Delahaye:2013jla}) or $e^+$  scattering on a target (LEMMA~\cite{Antonelli:2013mmk,Antonelli:2015nla,Collamati:2017jww}), see Ref.~\cite{Collamati:2017jww,Delahaye,wulzer-ALBA} for the attainable luminosities with these technologies.
Even more futuristic HELCs are linear electron colliders like AWAKE~\cite{Assmann:2014hva} and ALEGRO~\cite{Cros:2017jxp}, where the electrons are accelerated through proton-driven plasma wakefield acceleration PWFA (see for example Ref.~\cite{Adli:2013npa}, and Ref.~\cite{Delahaye} for the related luminosity).
We summarise in Figure~\ref{fig:lep_colliders} a selection of the different proposals in terms of their center-of-mass energy and their luminosity per year at the various stages.

Low energy $e^+e^-$ colliders are known to be wonderful machines to achieve very precise indirect measurements of the properties of the Higgs sector, well beyond the reach of HL-LHC (see Ref.~\cite{Craig:2017gzf} for a recent discussion).
As a concrete example FCC-ee could probe Higgs couplings to SM gauge bosons at the $0.1\%$ level, by collecting $10\text{ ab}^{-1}$ at the peak of Higgs-strahlung rate~\cite{Dawson:2013bba}.
In HELCs like CLIC the production cross-section for the SM Higgs will instead be dominated by the $WW$-fusion process, with a rate given in the high-energy limit by (see e.g.\ Ref.~\cite{Kilian:1995tr})
\begin{equation}\label{Wfusionxsec}
\sigma_{e\bar e \to h+\nu\bar\nu}= \frac{g^4}{256\pi^3}\frac{1}{v^2}\left[ \log \frac{s}{m_h^2} -2 +\mathcal{O}\Big(\frac{m_h^2}{s}\Big)\right],
\end{equation}
where $s$ is the center of mass energy, $v=246\ \GeV$ and $m_h \simeq 125.1$~GeV is the Higgs boson  mass. Since the $WW$-fusion rate depends only logarithmically on $s$, while the Higgs-strahlung rate is suppressed by $1/s$,  the cross-section in Eq.~\eqref{Wfusionxsec} gets larger than $\sigma(e\bar e \to Zh)$ for $\sqrt{s}\gtrsim400\text{ GeV}$. Therefore, by means of the $WW$-fusion process, HELCs can partially overcome the smaller luminosity compared to the low energy colliders and obtain a similar precision in the Higgs couplings to gauge bosons, as discussed in Ref.~\cite{CLIC-Higgs} (see also Ref.~\cite{Dawson:2013bba}). 

The purpose of this work is to assess the HELCs capabilities to discover new physics \textit{directly}, i.e. to produce and detect new particles, in particular if weakly coupled to the Standard Model and therefore at the edge of, or beyond, the reach of the HL-LHC.
In order to discuss the interplay between precision tests of the Higgs sector and the direct exploration of new physics, we focus on models where new physics is coupled to the SM Higgs sector.  To keep the discussion minimal, but retaining all the features discussed so far,  we consider a model where only a new real scalar singlet is added  to the SM. The new state mixes with the Higgs such that $i)$ it induces a tree-level shift in the Higgs couplings, $ii)$ it can be singly produced analogously to the Higgs, and $iii)$ it can be pair-produced via the quartic Higgs portal coupling.  Having set the framework, the main question we want to address in this paper is the following: 

\emph{To what extent will CLIC, and HELCs in general, directly test new physics inducing deviations in the Higgs couplings to gauge bosons of a few percent or smaller?}  

\medskip

A first answer to this question comes from the study of single production. From a full collider study, focusing on the dominant decays of the scalar singlet into di-Higgs and di-boson final states, we find that searches for resonances in four $b$-jets are going to be the dominant discovery/exclusion channel at CLIC.\footnote{See also Ref.~\cite{Chacko:2017xpd} for a detailed study of the search in the $4b$ final state at ILC, and an estimate for the 1.5 TeV CLIC.}
Interestingly, CLIC can test new resonances well beyond the capabilities of HL-LHC and down to couplings correlated to a deviation in the Higgs couplings smaller than 0.1$\%$.
We then carry out a similar analysis at muon colliders with center-of-mass energies of 6 or 14 TeV, like LEMMA or MAP, and find that direct searches for extra scalars would be more effective than the corresponding ones at a 100~TeV $pp$ collider.
We then recast our results in two explicit models featuring an extra scalar singlet, the NMSSM~\cite{Ellwanger:2009dp} and Twin Higgs~\cite{Chacko:2005pe}, that are well-motivated as they address the hierarchy problem of the Fermi scale. We also comment on to what extent di-boson searches at HELCs can probe the heavy mass regime of an axion-like particle that couples only to electroweak (EW) gauge bosons, a scenario which is notoriously challenging from the phenomenological point of view~\cite{Bauer:2017ris,Brivio:2017ije,Craig:2018kne}. 

A second answer comes from the study of pair production.
This becomes very important when the coupling controlling the single production is parametrically suppressed, as it happens when the new scalar is odd under an approximate $Z_2$ symmetry.
In this limit, pair production of the new scalars constitutes a direct test of the Higgs portal coupling.
Models of this type have been considered as interesting benchmarks for electroweak baryogenesis (see Ref.~\cite{Morrissey:2012db} for a review), because the coupling of the singlet with the Higgs can induce a first order electroweak phase transition (FOEWPT), see e.g.\ Ref.~\cite{Pietroni:1992in,Curtin:2014jma,Craig:2014lda}.
Depending on the decay length, the singlet pair production can lead to final states with multiple gauge or Higgs bosons, with displaced vertices in the tracker/muon chamber, or with missing energy.
We find that, in all of the cases above, CLIC (and a fortiori muon colliders) will sensibly ameliorate the HL-LHC reach and, most importantly, has the potential to entirely probe the region where a FOEWPT is possible.

\medskip

This paper is organised as follows. In Section~\ref{sec:setup} we introduce the model and the relevant formulae. Section~\ref{sec:single} is dedicated to the discussion of the collider study for single production, and  Section~\ref{sec:models} to its application to concrete models. In Section~\ref{sec:double} we present our findings for double production. We conclude in Section~\ref{sec:conclusions}.

\section{The Singlet Model}\label{sec:setup}

To set our notation, we consider here an extra real CP-even scalar degree of freedom coupled to the standard model only via renormalisable interactions
\be\label{modello}
\mathscr{L}= \mathscr{L}_{\rm SM} + \frac{1}{2}(\partial_\mu S)^2 - \frac{1}{2}m_S^2 S^2 - a_{HS} S |H|^2 - \frac{\lambda_{HS}}{2} S^2 |H|^2 -\frac{a_{S}}{3} S^3 -\frac{\lambda_S}{4} S^4,\quad \quad H=\left(\!\!\!\begin{array}{c}\pi^+\\ \frac{v+h_0+i\pi_0}{\sqrt{2}}\end{array}\!\!\!\right).
\ee
We define the mixing angle $\gamma$ as the rotation needed to go from the basis of Eq.~\eqref{modello}, where only the Higgs couples directly to the SM fields, to the mass basis
\begin{align}
h &= h_0 \cos\gamma  + S \sin\gamma, & \phi &= S \cos\gamma - h_0 \sin\gamma,\label{eq:massbasis}
\end{align}
where $h$ is the SM-like Higgs with mass $m_h = 125$ GeV, and $\phi$ is the singlet-like state with mass $m_\phi$. 

We now highlight the main phenomenological consequences of the Lagrangian in Eq.~\eqref{modello}, discussing both deviations in the SM-like Higgs couplings, and single and double production of the new scalar $\phi$. For definiteness, in this paper we only consider the mass ordering $m_\phi > m_h$.

\paragraph{SM-like Higgs boson.}
The main deviation in the Higgs couplings to vectors and fermions is generated at tree-level by the mixing $\gamma$: the Higgs signal strengths $\mu_h$ are universally rescaled as 
\be
\mu_h =\mu_h^{\rm SM} \cos^2\gamma\,.
\label{Higgs_signal_strengths}
\ee
When $a_{HS}$ and $\lambda_{HS}$ are both non-vanishing, the above deviations are uncorrelated from those in the trilinear Higgs coupling, that can in principle be larger.
Under favourable circumstances, the HL-LHC could even observe deviations in double Higgs production without observing any in the Higgs couplings to SM fields, see e.g.\ Ref.~\cite{Buttazzo:2015bka}.

An accurate description of the Higgs sector in our setup can also be achieved by integrating out the singlet field and computing the Wilson coefficients of the dimension-6 operators
\be
\mathscr{L}=\mathscr{L}_{\rm SM} + \frac{c_H}{\Lambda^2} O_H - \frac{c_6\lambda_H}{\Lambda^2} O_6\, ,\qquad\quad\text{with} \quad O_H = \frac{1}{2}(\partial_\mu |H|^2)^2, \quad O_6 = |H|^6\, ,
\ee
where $\lambda_H$ is defined as the coefficient of the $|H|^4$ operator.
These operators predict the following relative shifts in the Higgs couplings 
\be
\frac{g_{hVV,ff}}{ g_{hVV,ff}^{\rm SM} }= 1 -\frac{c_H}{2} \frac{v^2}{\Lambda^2} + \mathcal{O}\Big(\frac{v^4}{\Lambda^4}\Big),\qquad\qquad
\frac{g_{hhh}}{g_{hhh}^{\rm SM}}=1+\big(c_6-\frac{3}{2}c_H\big)\frac{v^2}{\Lambda^2}+ \mathcal{O}\Big(\frac{v^4}{\Lambda^4}\Big)\,, \label{eq:Higgscoupling}
\ee
where $v=246~\GeV$ and $g_{hhh}^{\rm SM} = 3 m_h^2/2v$ (see also Ref.~\cite{Contino:2013gna} for a related discussion at lepton colliders).
In the singlet model, the tree level and one-loop contributions to these operators read 
\be\label{matching}
\frac{c_H}{\Lambda^2}=\frac{\sin^2\gamma}{ v^2}+\frac{\lambda_{HS}^2}{192\pi^2 m_\phi^2}\,,\qquad\qquad \frac{c_6\,\lambda_H}{\Lambda^2}= \frac{\lambda_{HS}}{2v^2}\sin^2\gamma + \frac{\lambda_{HS}^3}{192\pi^2 m_\phi^2}\,.
\ee
Notice that we assume negligible cubic terms for the singlet to compute the first contribution to $c_6$ (see Ref.~\cite{Henning:2014wua,Cao:2017oez}).
This shows the importance of $\sin^2\gamma$ in the low-energy phenomenology of the Higgs, and the possible interplay between a small mixing and a large portal coupling $\lambda_{HS}$ to get visible effects in the triple Higgs coupling.

\paragraph{Singlet-like $\phi$ boson.}
The phenomenology of $\phi$ depends largely on the presence of a non-vanishing mixing angle (i.e. of a breaking of the $Z_2$ symmetry), given that in this case it can be singly produced, and it can decay to SM particles. Single production cross-sections and SM decay rates for the singlet-like state $\phi$ are proportional to $\sin^2\gamma$, and read
\begin{align}
\sigma_{\phi} &= \sin^2\gamma\cdot \sigma_{h}(m_\phi),\label{sigmaphi}\\
{\rm BR}_{\phi\to f\bar f, VV} &= {\rm BR}_{h\to f\bar f, VV} (1 - {\rm BR}_{\phi\to hh}),\label{BRphi}
\end{align}
where $\sigma_h(m_\phi)$ is the production cross-section for a SM Higgs boson of mass $m_\phi$ and in the second equation we assumed the absence of non-SM decay modes.  From Eq.~\eqref{BRphi} we see that the heavy singlet dominantly decays into $W$, $Z$, and Higgs bosons. The branching ratio into $hh$ is in principle a free quantity that depends on the parameters of the scalar potential, but for heavy scalars $m_\phi \gg m_W$ the potential exhibits an approximate SO(4) symmetry which implies ${\rm BR}_{\phi\to hh} \simeq {\rm BR}_{\phi\to ZZ} \simeq {\rm BR}_{\phi\to WW}/2$.

Double production of the singlet can instead proceed at tree-level even for $\gamma=0$, through the portal coupling $\lambda_{HS}$. We show in Section~\ref{sec:formule} the explicit dependence of the production rates on the masses and couplings of the model.

\subsection{Scaling of physical quantities}\label{sec:scaling}

The actual size of the parameters appearing in Eq.~\eqref{modello} depends on the concrete UV model under consideration, in such a way that measuring single and/or double production could provide a hint about the underlying dynamics. In order to maintain a simple description, before specialising to concrete cases in Section~\ref{sec:models}, we consider a new sector with two intrinsic parameters: a mass scale $M_*$ and a coupling $g_*$ (see also Ref.~\cite{Chala:2017sjk}). In this case we can write 
\begin{align}
&m_S^2 \approx M^2_*, \qquad \lambda_{HS} \approx g_*^2, \qquad  \lambda_S \approx g_*^2 \qquad a_{HS} \approx g_* M_*,\qquad a_S \approx g_* M_*\ ,
\end{align}
where the proportionality constants are numerical coefficients of O(1), unless constrained by additional symmetries and/or scales. The squared mass matrix of the two states $(h_0,S)$ reads 

\be
\mathcal{M} = \left(\begin{array}{cc}2\lambda_H v^2 & v(a_{HS} + \lambda_{HS} s) \\  v(a_{HS} + \lambda_{HS} s) & m_S^2 + 2 a_S s + 3 \lambda_S s^2 + \lambda_{HS} v^2/2 \end{array}\right)\,,
\ee
where $s$ is the vacuum expectation value (VEV) of the singlet. Under the assumption of a small mixing angle as suggested by data, $\gamma \simeq \mathcal{M}_{12}/\mathcal{M}_{22}$.
The size of $\gamma$ depends on the origin of the singlet VEV $s$. We can identify three different scenarios:
\begin{enumerate}
\item If the singlet develops a VEV because of its potential alone, then $s\approx M_*/g_*$, so that the mixing angle scales as
\be\label{linearscaling}
\gamma \approx \frac{g_* v}{M_*}\,,
\ee
times a combination of $O(1)$ numbers (that can be tuned to be small).
\item If the singlet gets its VEV due to the interaction with the Higgs dynamics, then $s\approx a_{HS} v^2/g_*^2 M_*^2$. In this case, the mixing angle is controlled by the size of the explicit breaking of the $S\to -S$ symmetry,
\be\label{Z2scaling}
\gamma\approx \frac{v a_{HS}}{M_*^2} = \frac{g_* v}{M_*}\delta,
\ee
where the dimensionless parameter $\delta = a_{HS}/g_* M_*$ can be made arbitrarily small.
\item If, in addition to the previous case, the dynamics responsible for the $Z_2$ breaking terms is related to a mass scale independent of $M_*$, then the mixing decouples as $M_*^{-2}$ and not as $M_*^{-1}$.
\end{enumerate}
These situations are realised in different concrete examples discussed in the literature:
\begin{itemize}
\item[$\diamond$] Models with a moderate coupling $g_*$ with $a_{HS} = 0$ and $\lambda_{HS} \sim g_*^2$. In this case we expect a deviation $\gamma\approx g_* v/M_*$, so that the only way to comply with the bounds is by $M_*/g_*\equiv f \gg v$. This type of decoupling is a well known feature of Twin Higgs models (see Section~\ref{sec:TH}).
\item[$\diamond$] Models with a weak coupling $g_*$, such that $\gamma\approx g_* v/M_*\ll 1$ even for light states. This is the case in the NMSSM, once we identify $M_*$ with the SUSY-breaking mass of the singlet ($M_*=\tilde m$) and $g_\ast$ with the coupling in superpotential ($g_\ast=\lambda$). The only way to additionally suppress the mixing angle is to invoke a tiny $s$, which is achievable by neglecting the $A_\kappa S^3 +h.c.$ soft term and by allowing $a_{HS} \equiv A_\lambda \sin(2\beta) \ll \lambda \tilde m$ (see Section \ref{sec:NMSSM}).
\end{itemize}
Notice also that the bounds obtained at the kinematic edge of the lepton collider, where $\gamma$ quickly approaches O(1) for large masses, could be interpreted in terms of strongly coupled new physics. This region however is (and will be) strongly constrained by single Higgs production.

\subsection{Vector boson fusion}\label{sec:formule}

\begin{figure}[t]
\centering
\includegraphics[width=.45\textwidth]{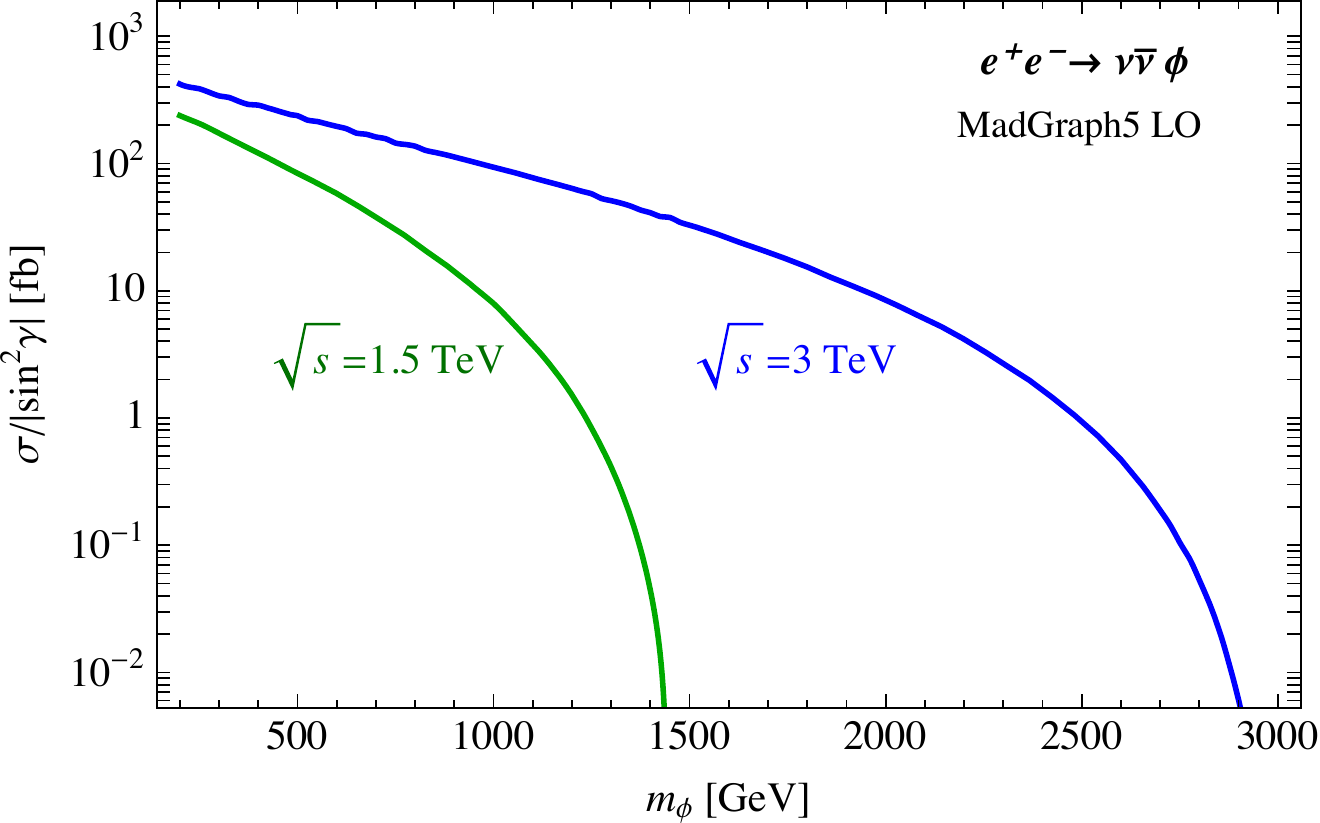}\hfill%
\includegraphics[width=.45\textwidth]{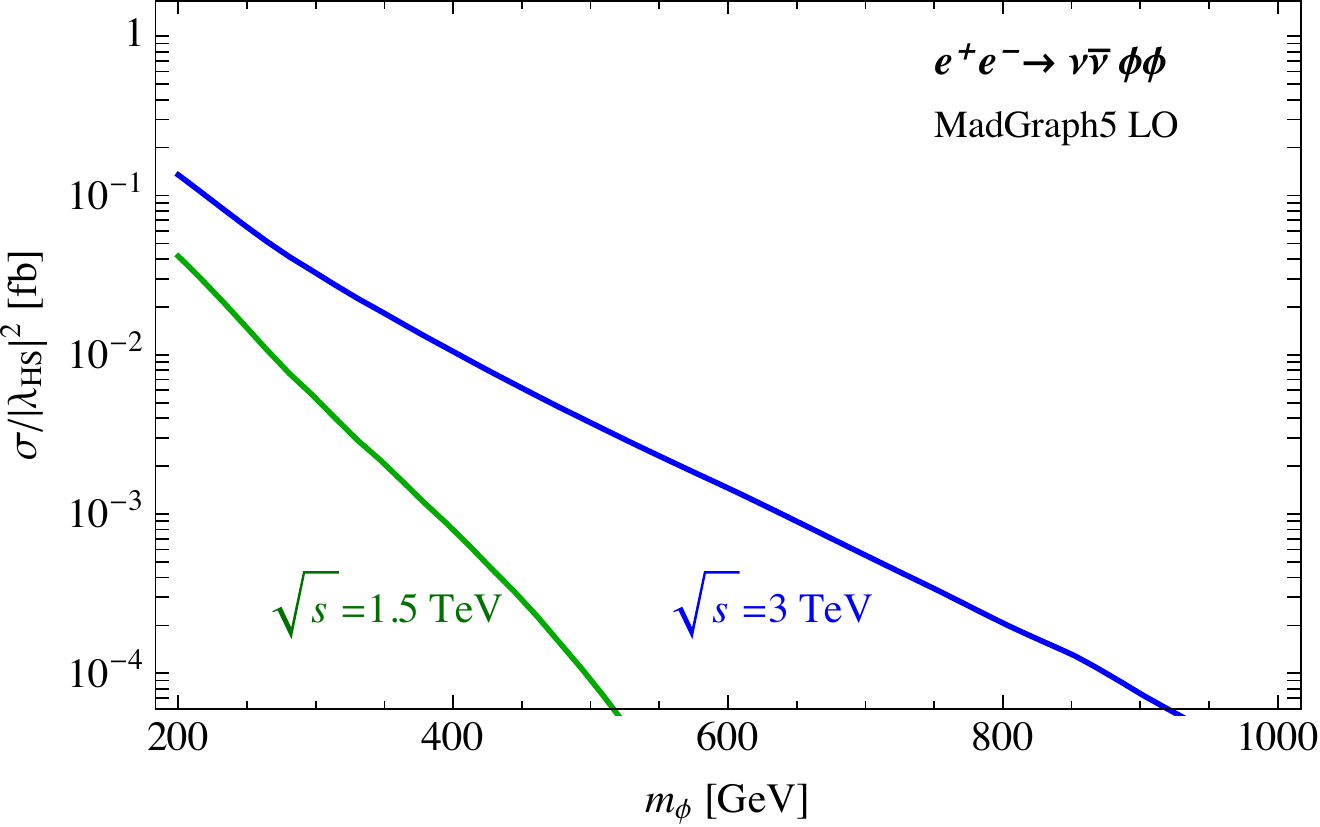} 
\caption{ {\bf Left:} single production via $WW$-fusion of a singlet. {\bf Right:} pair production induced via $WW$-fusion of singlets, assuming $\sin^2\gamma=0$.}
\label{fig:rate-theory}
\end{figure}

As discussed in the introduction, the advantage of HELCs is mainly due to the effectiveness of vector boson fusion as a production mode for scalar particles. Both single and double productions can be written in terms of the cross-section of the subprocess $VV\to \phi$ and $VV\to \phi\phi$ properly convoluted with the splitting functions for $\ell \to V \ell'$. Any differential distribution for the process $e \bar e\to \nu\bar\nu X$ can be written as a distribution in the invariant mass squared of the subprocesses as
\be
\frac{d\sigma}{d\hat{s}}=\frac{\hat \sigma_{V_iV_j\to X}(\hat{s})}{s} \, \mathscr{C}_{V_i V_j}(\hat s),\quad \mathrm{with}\quad \mathscr{C}_{V_i V_j}(\hat s)=\int_{\hat s/s}^1 \frac{dx}{x} f_{V_i}(x)f_{V_j}(\frac{\hat s x}{s})\,,
\ee
where we defined the effective parton luminosities $\mathscr{C}_{V_iV_j}$ in terms of the splitting functions $f_{V_i}(x)$. These can be computed analytically in the regime $M_V^2/\hat s \ll 1$ \cite{Chanowitz:1985hj,Dawson:1984gx}. Here we focus on the longitudinal polarisations, which are the only ones coupled to the extra singlet through the mixing with the SM Higgs:
\be
f_{W_L}(x)\simeq\frac{g^2}{64\pi^2}\frac{1-x}{x}\, ,\qquad \mathscr{C}_{W_L W_L}(\hat s)=\frac{g^4}{4096\pi^4} \frac{s}{\hat s} \big[ (1+\frac{\hat s}{s}) \log \frac{s}{\hat s} + 2(\frac{\hat s}{s}-1)\big]\, .
\ee
By inspecting the behaviour at high $s$, we see that the total rate of $WW$-fusion does not fall with energy neither for single nor for double singlet production. The total rates can be computed to be
\begin{align}
\label{single}
&\sigma_{e\bar e \to\nu\bar \nu S}= \sin^2\gamma\, \frac{g^4}{256 \pi^3}\frac{1}{v^2} \left[2 \Big(\frac{m_\phi^2}{s}-1\Big)+\Big(\dfrac{m_\phi^2}{s}+1\Big) \log\frac{s}{m_\phi^2}\right]\simeq \sin^2\gamma\frac{g^4}{256\pi^3} \frac{\log\frac{s}{m_\phi^2}-2}{v^2} ,\\ 
\label{double}
&\sigma_{e\bar e \to \nu\bar \nu SS}= \frac{g^4|\lambda_{HS}|^2}{49152 \pi^5}\,\frac{1}{m_\phi^2} \Big[\log \frac{s}{m_\phi^2}- \dfrac{14}{3} + \frac{m_\phi^2}{s} \big(3 \log^2\frac{s}{m_\phi^2} + 18 -\pi^2\big) + \mathcal{O}\Big(\frac{m_\phi^4}{s^2}\Big)\Big],
\end{align}
where Eq.~\eqref{double} holds in the limit $\sin\gamma = 0$.
The formulas in Eq.~\eqref{single}--\eqref{double} are extremely good approximations  as long as the dominant contribution to the rates comes from kinematic configurations where $M_V^2/\hat s \ll 1$. We checked that they reproduce with excellent accuracy the full result, which we compute with \textsc{MadGraph5}~\cite{Alwall:2011uj,Alwall:2014hca}. This is shown in Figure~\ref{fig:rate-theory} and we use it in all our numerical calculations. Here and in what follows, we assume unpolarised electron beams.

The above expressions for the production rate show explicitly what is well known: $WW$-fusion is a powerful production channel for HELCs.
At increased center-of-mass energy, other production mechanisms such as $\phi$-strahlung and double $\phi$-strahlung are subdominant, because they are suppressed at large $s$ (see also Ref.~\cite{Kilian:1995tr} for a comparison).
Based on these considerations we motivate our approach of just considering $VV$-fusion processes for the production of the scalar singlet. 
In our study we do not include next-to-leading orders in EW radiation, thus an uncertainty of the order of $\frac{\alpha_2}{4\pi}\log(s/m_W^2)$ should be understood in all our sensitivities.\footnote{The production of the new singlet is driven by its couplings to the longitudinal components of SM vectors thus it has only one logarithm from the collinear singularity. This is not true for the background, but its impact on the uncertainty of the sensitivities would be subleading because it is dominated by statistics. See also Ref.~\cite{Chen:2016wkt}.} This is safely below the 3\% level even at the 14 TeV stage of future $\mu$-colliders.

\section{Single production}\label{sec:single}

In this section we assess the capabilities of HELCs to test the existence of new scalar particles by means of their single production in W-fusion. The total production rate as a function of the mass of the scalar has been computed in the previous section, and is displayed in the left panel of Figure~\ref{fig:rate-theory}. The dominant decay channels of $\phi$ are into pairs of vector bosons and Higgs bosons, as given in Eq.~\eqref{BRphi}. We are going to study resonant production modes, in narrow-width approximation and with only visible final states, and thus we perform our analyses in the ``cut-and-count" scheme. The significance of a given number of signal events $N_{\rm sig}$ around the resonance peak, against a background $N_{\rm bkg}$, is defined as
\be\label{eq:significance}
\mathrm{significance}=\frac{N_{\rm sig}}{\sqrt{(N_{\rm sig}+N_{\rm bkg})+\alpha_{\rm sys}^2 N_{\rm bkg}^2}}\,,
\ee
where $\alpha_{\rm sys}$ are the systematic and theoretical uncertainties on the SM rates.
For definiteness, in what follows we always set $\alpha_{\rm sys} = 2\%$.
As we will show, all our results are dominated by statistics up to systematic errors of 10\% or larger.
We refer to Appendix~\ref{app:lepton} for a precise assessment of the impact of different choices for $\alpha_{\rm sys}$.

\medskip

Before entering into the details of the analysis,
to set a reference for the sensitivities, we compute the best possible reach that one would achieve in the case of negligible background. We define it as the signal cross section that results in 3 signal events
\be\label{0background}
\sigma(e^+e^- \to \phi \nu\bar\nu)\times {\rm BR}(\phi\to f) \simeq 3/L,
\ee
where $L$ is the integrated luminosity. Using Eq.~\eqref{single}, this limit translates into an approximate sensitivity on the mixing angle 
\be\label{saturated}
\sin^2\gamma \times {\rm BR}(\phi\to f) \approx 0.02 \left( \frac{1/\fb}{L} \right) \times \left[\log\frac{s}{m_\phi^2} - 2 + \frac{m_\phi^2}{s}\Big(\log\frac{s}{m_\phi^2} +2\Big)\right]^{-1}\,.
\ee
Notice the logarithmic dependence on the particle mass for $m_W^2\ll m_\phi^2 \ll s$, explaining why our sensitivities are almost flat when compared with those obtained at hadron colliders. The aim of the following two sections is to determine how much a realistic analysis can approach the sensitivity in Eq.~\eqref{saturated}. 

We now discuss the reach at different center-of-mass energies in the dominant decay channels $hh$, $ZZ$, and $WW$. As we show below, the sensitivities from the $hh(4b)$ decay mode turn out to be very strong at lepton colliders. For this reason we start performing a detailed simulation of this channel, while we simply work at parton level (before showering) for the leptonic and semi-leptonic $VV$ decays.

\subsection{Decay channel $\phi \to hh$}\label{sec:Stohh}

\begin{figure}[t!]
\centering%
\includegraphics[width=0.5\textwidth]{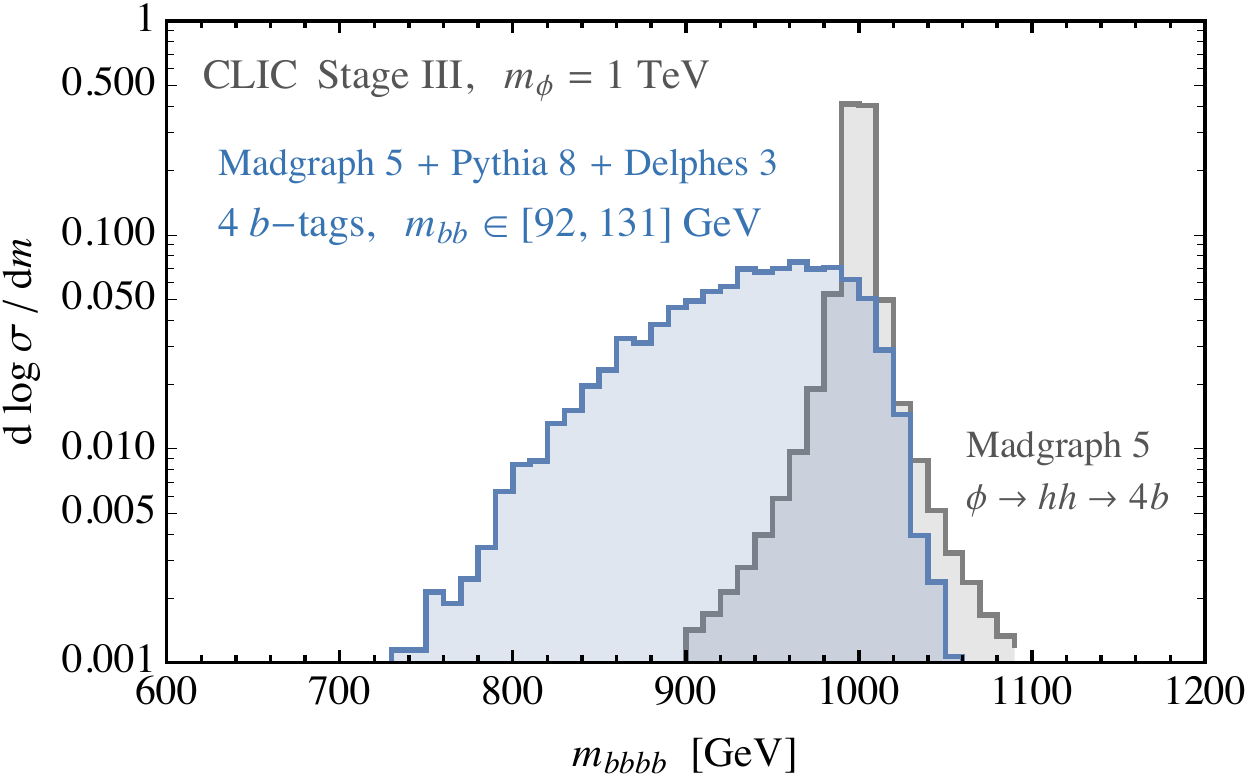}\hfill%
\includegraphics[width=0.48\textwidth]{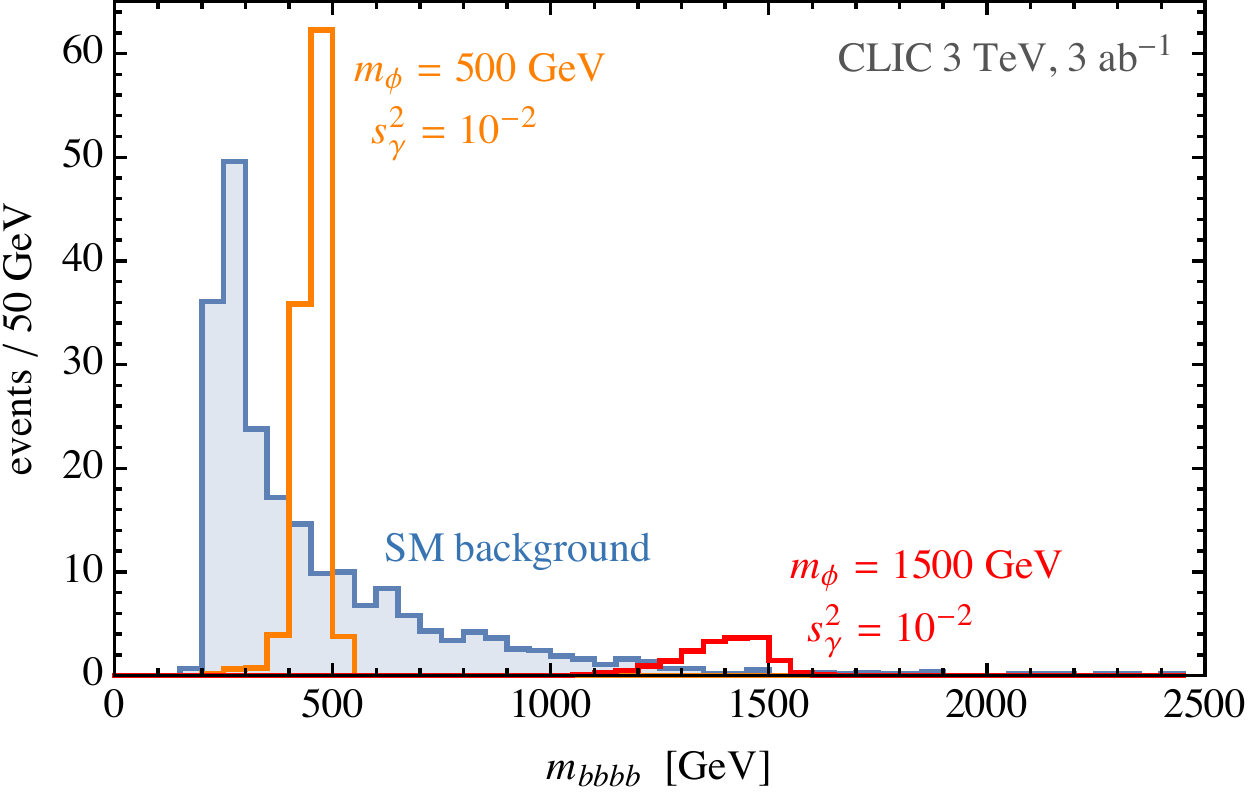}
\caption{{\bf Left:} invariant mass distribution of the 4 $b$-quarks in the signal, for $m_\phi = 1$ TeV at the 3 TeV CLIC. The blue histogram shows the signal after parton showering, detector simulation, and identification cuts; the grey line shows the output of the Monte Carlo generator before parton showering. {\bf Right:} $4b$ invariant mass distribution of the SM background, with two examples of signal superimposed.  \label{fig:invmass}}
\end{figure}

In the model under consideration the largest individual branching fraction of the singlet is $\phi\to hh(4b)$. We look for this signal as a narrow resonant contribution over the SM background in the $4b$ invariant mass distribution. The same signature has been studied in \cite{Chacko:2017xpd}, where the authors discuss the reach of ILC and CLIC 1.5 TeV. With respect to that work we include a full CLIC detector simulation.

Requiring $W$-fusion production, the principal background is the irreducible SM contribution to $e^+ e^-\to 2\nu 4b$, with a dominant component due to $hh(4b)$ and $Zh(4b)$. The total cross-section for this process is computed with {\sc MadGraph} to be 1.8 fb (0.6 fb) at the center-of-mass energy of 3 TeV (1.5~TeV). A potentially large reducible contribution from $\gamma\gamma \to 4b$ is avoided imposing cuts on the transverse momentum of the $b$ quarks ($p_{T} > 20\, {\rm GeV}$) and on the missing energy ($E_{\rm miss}~>~30~{\rm GeV}$), and turns out to be completely negligible. 

We also compute the cross-sections for the signal $e^+e^- \to \phi(4b) \nu\bar\nu$ with {\sc MadGraph}, after implementing the Lagrangian in Eq.~\eqref{modello} in {\sc FeynRules} 2.0~\cite{Alloul:2013bka}, always working in the narrow-width approximation for the singlet, and retaining the subdominant contribution from $\phi\to ZZ$. We use \textsc{Pythia8}~\cite{Pythia8} for showering and \textsc{Delphes3}~\cite{deFavereau:2013fsa} for detector simulation, using the configuration of the CLIC cards of Ref.~\cite{ulrike}. We apply the VLC exclusive jet reconstruction algorithm \cite{Boronat:2014hva} with working point $R=0.7$ and $N=4$ (see also Ref.~\cite{Boronat:2016tgd}): this allows us to reconstruct $b$-jets with $\Delta R$ as small as about 0.1, well below the standard isolation cut, compatibly with the detector resolution expected at CLIC (see Appendix~\ref{app:4b} for more details).

\begin{table}[t]
\small
\renewcommand{\arraystretch}{1.15}
\begin{subtable}{0.5\textwidth}
\centering%
\begin{tabular}{ccc}
\hline
Cut & $\epsilon_{\rm sig}$ & $\epsilon_{\rm bkg}^{4b2\nu}$\\
\hline
$E_{\rm miss} > 30$ GeV & 90\% & 95\%\\
4 $b$-tags & 50\% & 35\%\\
$m_{bb} \in [88,129]$ GeV & 64\% & 23\%\\
$|\cos\theta| < 0.94$ & 96\% & 63\%\\
$m_{4b} \in [770,1070]$ GeV & 98\% & 2.8\%\\
\hline
Total efficiency & 27\% & $1.3\times 10^{-3}$\\
\hline
\end{tabular}
\caption{CLIC 1.5 TeV, $m_\phi = 1$ TeV}
\end{subtable}\hfill%
\begin{subtable}{0.5\textwidth}
\centering%
\begin{tabular}{ccc}
\hline
Cut & $\epsilon_{\rm sig}$ & $\epsilon_{\rm bkg}^{4b2\nu}$\\
\hline
$E_{\rm miss} > 30$ GeV & 94\% & 96\%\\
4 $b$-tags & 51\% & 33\%\\
$m_{bb} \in [88,137]$ GeV & 60\% & 15\%\\
$|\cos\theta| < 0.95$ & 97\% & 58\%\\
$m_{4b} \in [1.5,2.04]$ TeV & 91\% & 0.7\%\\
\hline
Total efficiency & 26\% & $2 \times 10^{-4}$\\
\hline
\end{tabular}
\caption{CLIC 3 TeV, $m_\phi = 2$ TeV}
\end{subtable}
\caption{Efficiencies for signal and background in $e^+e^-\to 4b\, 2\nu$, for each individual cut applied in the analysis. The two cases $m_\phi = 1$ TeV and $m_\phi = 2$ TeV are shown, respectively, for CLIC Stage II and Stage III.\label{Tab:cutflow}}
\end{table}

In order to select the events we proceed with the following steps:
\begin{enumerate}
\item We impose a cut on the transverse momentum of the jets $p_T > 20$ GeV and on the missing energy $E_{\rm miss} > 30$ GeV in order to select events coming from $W$-fusion.
\item  $b$-tagging: we require the presence of four jets tagged as $b$, using the loose selection criterion as implemented in Ref.~\cite{ulrike} in order not to excessively reduce the signal efficiency.
\item $h$ reconstruction: we identify the candidate Higgs bosons by choosing the pairing of the four $b$-jets that gives reconstructed invariant masses of the two Higgses closest to 125 GeV, i.e.\ the one that minimises the quantity $(m_{b_1 b_2} - 125\,\GeV)^2+(m_{b_3 b_4} - 125\,\GeV)^2$. We then retain the events having two distinct $b$-pairs with $m_{b\bar b}$ in a window of about [90, 130] GeV. The exact boundaries of the invariant-mass window are chosen differently for each $m_\phi$ hypothesis, in order to maximise the significance of the signal.
\item We apply a cut on the polar angle $\vert\cos\theta\vert \lesssim 0.9$ of the two Higgs bosons, in order to reduce the contribution from the forward region, where the background is enhanced. The precise value of the cut is chosen for each value of the mass in order to maximise the significance.
\item $\phi$ reconstruction: we select the events with a total invariant mass of the $4b$ system in a window of about $0.75\, m_\phi \lesssim m_{4b} \lesssim 1.05\, m_\phi$ around the resonance peak, again optimising the cut for each signal hypothesis.
\end{enumerate}
Figure~\ref{fig:invmass} (left) shows the invariant-mass distribution of the 4 $b$ quarks for the signal, comparing the result of the detector simulation, including $b$ and $h$ identification cuts, with the output of the Monte Carlo generator for $\phi\to hh(4b)$ before parton showering.
The efficiencies $\epsilon_{{\rm sig},{\rm bkg}}$ for signal and background of each step of the cut-flow are given in Table~\ref{Tab:cutflow} for two benchmark cases. We verified that these numbers do not vary substantially changing the $R$ parameter of the jet reconstruction algorithm, and changing the exact values of the kinematical cuts.
For the signal, the most important effects come from $b$-tagging and from the Higgs mass reconstruction, which both have efficiencies in the 40\%--60\% range, depending on the resonance mass and on the collider energy,\footnote{Notice that the rather low Higgs reconstruction efficiency for the signal is defined including the $\phi\to ZZ$ contribution.} for a total signal reconstruction efficiency of about 25\%. The SM background, on the other hand, is reduced by a factor of at least a few $10^{-3}$, reaching up to $10^{-4}$ for masses above a TeV at CLIC Stage III. In Figure~\ref{fig:invmass} (right) we show the $4b$ invariant-mass distribution for the background at CLIC Stage III, with two signal examples superimposed.
Notice that for singlet masses above a TeV the search becomes essentially background-free, and, after taking into account the efficiency of the signal, the limit roughly corresponds to the estimate in Eq.~\eqref{0background}.

The exclusion limits are computed from Eq.~\eqref{eq:significance} requiring a significance of 1.64, which corresponds to 95\% C.L. (one sided).
In Figure~\ref{CLIClimits} we show the results for the CLIC sensitivity in $\sigma(e^+e^- \to \phi \nu\bar\nu) \times {\rm BR}(\phi\to hh)$ as a function of the mass of the singlet, and compare it with the reach in the various $\phi\to VV$ channels described in Section~\ref{sec:StoVV}.
One can see that, despite the rather strong assumptions made to derive the other limits (see next section), the $4b$ channel turns out to be the best probe of scalar singlet production, if one assumes similar branching ratios into $hh$ and $ZZ$.
It is also evident that, due to the low number of background events over a large range of masses, the reaches depend only mildly on the collider energy and resonance mass as expected.

\subsection{Decay channels $\phi \to VV$}\label{sec:StoVV}

\begin{figure}[t!]
\centering
\includegraphics[width=.48\textwidth]{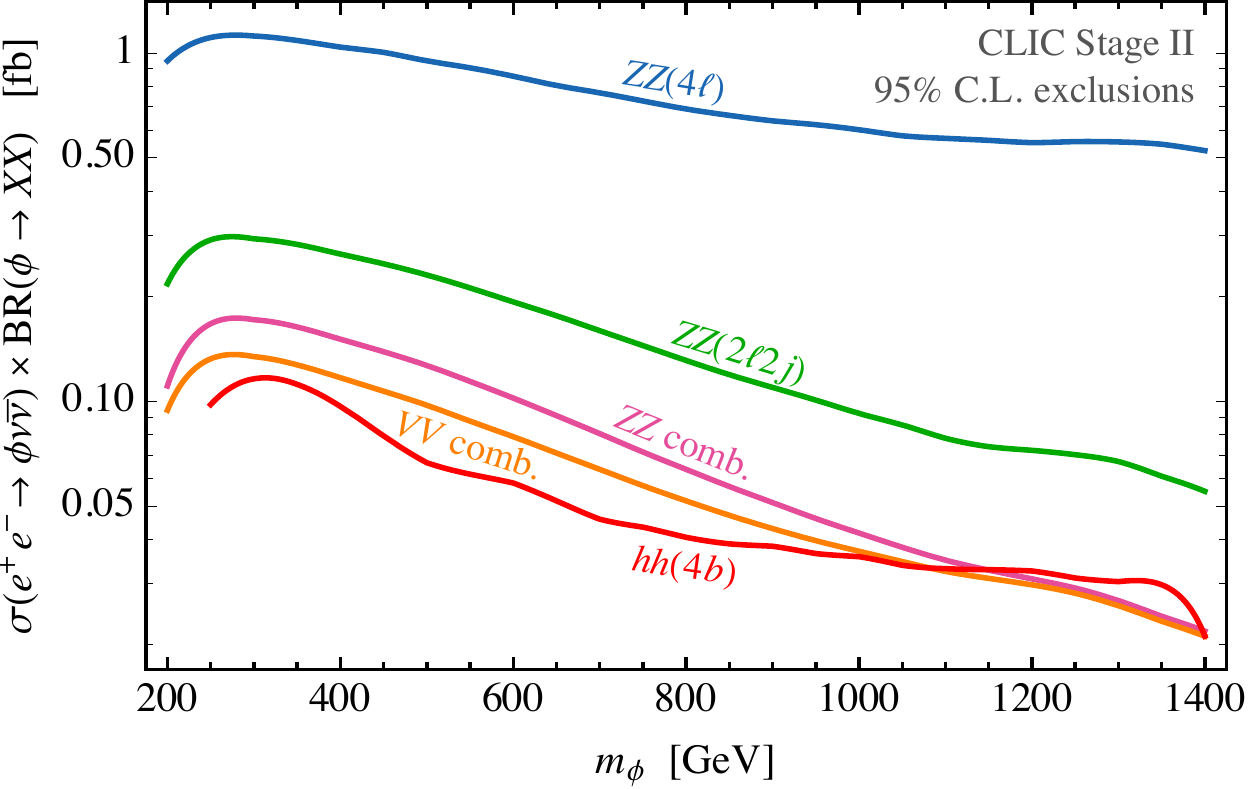}\hfill%
\includegraphics[width=.48\textwidth]{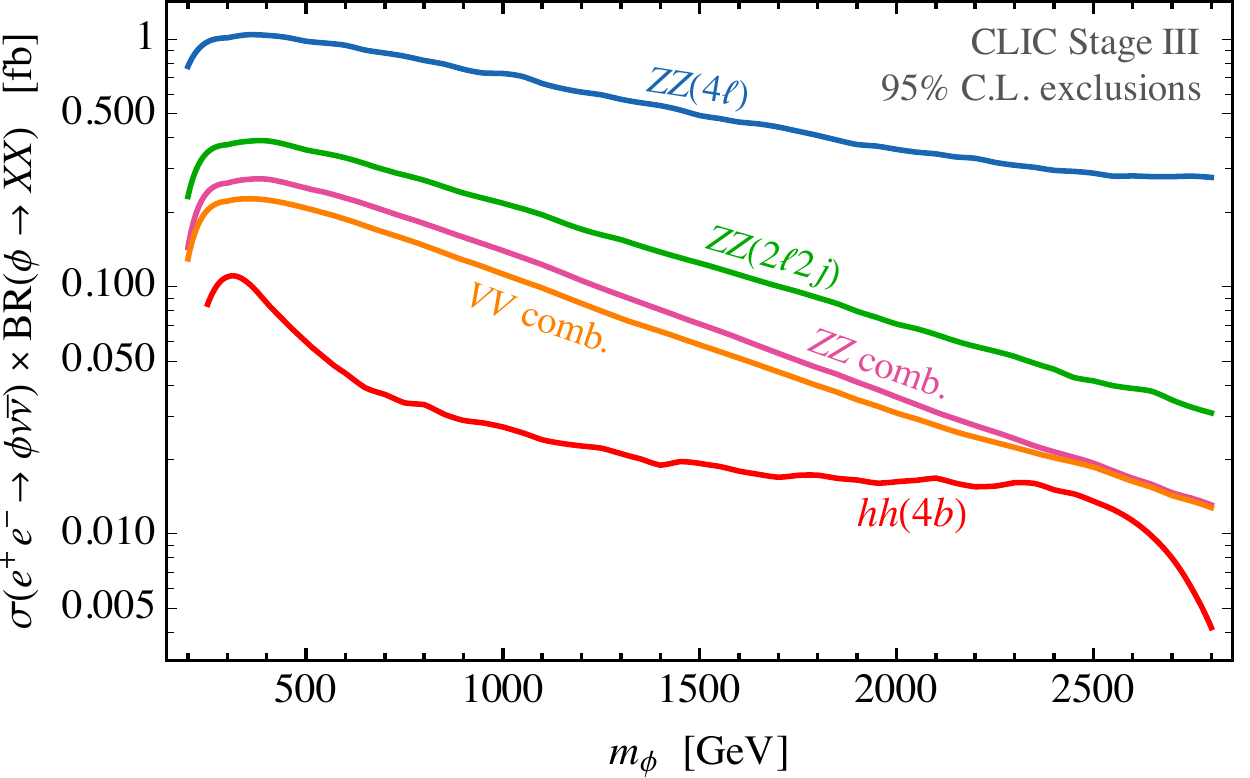} 
\caption{\label{CLIClimits}Comparison between the projected 95\% C.L. exclusions in different channels on $\sigma(e^+e^- \to \phi \nu\bar\nu)\times {\rm BR}(\phi\to XX)$, with $X = h, Z$. The limits, from top to bottom, come from $\phi\to ZZ\to 4\ell$ (blue), $\phi\to ZZ\to 2\ell 2j$ (green), combination of $\phi\to ZZ\to 4\ell, 2\ell 2j, 4j$ (pink), combination of $\phi\to ZZ$ and $\phi\to WW$ (orange, assuming ${\rm BR}_{\phi\to WW} = 2\, {\rm BR}_{\phi\to ZZ}$), and finally $\phi\to hh\to 4b$ (red). {\bf Left:} CLIC with $\sqrt{s} = 1.5$ TeV, $\mathcal{L} = 1.5\, {\rm ab}^{-1}$. {\bf Right:} CLIC with $\sqrt{s} = 3$ TeV, $\mathcal{L} = 3\, {\rm ab}^{-1}$.}
\end{figure}

Given the SM-like properties of the high-mass state $\phi$, a large fraction of singlets is expected to decay into pairs of vector bosons. 
We consider four kinds of resonant signals: 
$$\phi\to ZZ\to 4\ell, \qquad \phi\to ZZ\to 2\ell 2j, \qquad \phi\to ZZ\to 4j, \qquad  \phi\to WW\to 4j\,,$$
all of which allow a rather precise reconstruction of the resonance mass. Because of the excellent resolution expected at future lepton colliders, we do not perform a complete detector simulation for these channels, but we rather give an estimate of the reach obtained from the backgrounds calculated at parton level. The limits derived here thus do not stand on the same grounds as the ones of the previous section, and should be understood as optimistic estimates.

Considering again $W$-fusion production mode, the dominant backgrounds to the $e^+e^-\to\nu\bar{\nu}\phi(VV)$ signal originate from electroweak di-boson production. The cross-section rates for these processes, computed with {\sc madgraph}, are
\be\label{VVirreducible}
\sigma(e^+e^- \to \nu\bar{\nu}ZZ)_{3\,\TeV}=57\ \fb,\quad\quad \sigma(e^+e^- \to \nu\bar{\nu}WW)_{3\,\TeV}= 131\, \fb,
\ee
while the corresponding values at $\sqrt{s}=1.5\,\TeV$ are 18 fb and 52 fb.

In the $ZZ(4\ell)$ channel, the other main source of background is $e^+e^- \to e^- \nu W^+ Z$ (and its conjugate), in particular from tri-boson production, which however is reduced to 0.3 fb requiring the mass of the leptons to be within 5\% of $m_Z$. The same process, with a hadronically decaying $Z$, contributes also to $ZZ(2\ell2\nu)$. A few more words deserve to be spent for the hadronic modes. Given the broad shape of the hadronically decaying vectors, a non-negligible misidentification probability for $Z$ and $W$ should be taken into account in a realistic analysis. As a consequence, a contamination between the various $VV$ backgrounds could become important.
The dominant effect comes from the process $e^- \gamma \to \nu W^- Z$ and its conjugate, which is a background for both the $2\ell2j$ and the $4j$ channels and has a large cross-section of 163~fb at 3~TeV.
This contribution could however be suppressed by requiring sufficiently hard momenta for the vectors in the final state (see e.g.\ Ref.~\cite{Contino:2013gna}) and, for the $ZZ$ signal, by a sufficiently hard cut on the dijet invariant mass. We leave a detailed study of this background to a future work and, in the spirit of giving an aggressive estimate of the sensitivities, we ignore it in what follows.
The contamination between the $WW$ and $ZZ$ channels, on the other hand, requires a double misidentification, and is expected to be small. Therefore, in the following we assume that all $W(jj)$ and $Z(jj)$ will be told apart thanks to the excellent jet mass resolution at CLIC. Finally, we also assume that all backgrounds without neutrinos in the final state will become negligible after a suitable missing energy cut is imposed, and consider only the irreducible backgrounds of Eq.~\eqref{VVirreducible} for our analysis.

\begin{figure}[t!]
\centering%
\includegraphics[width=0.7\textwidth]{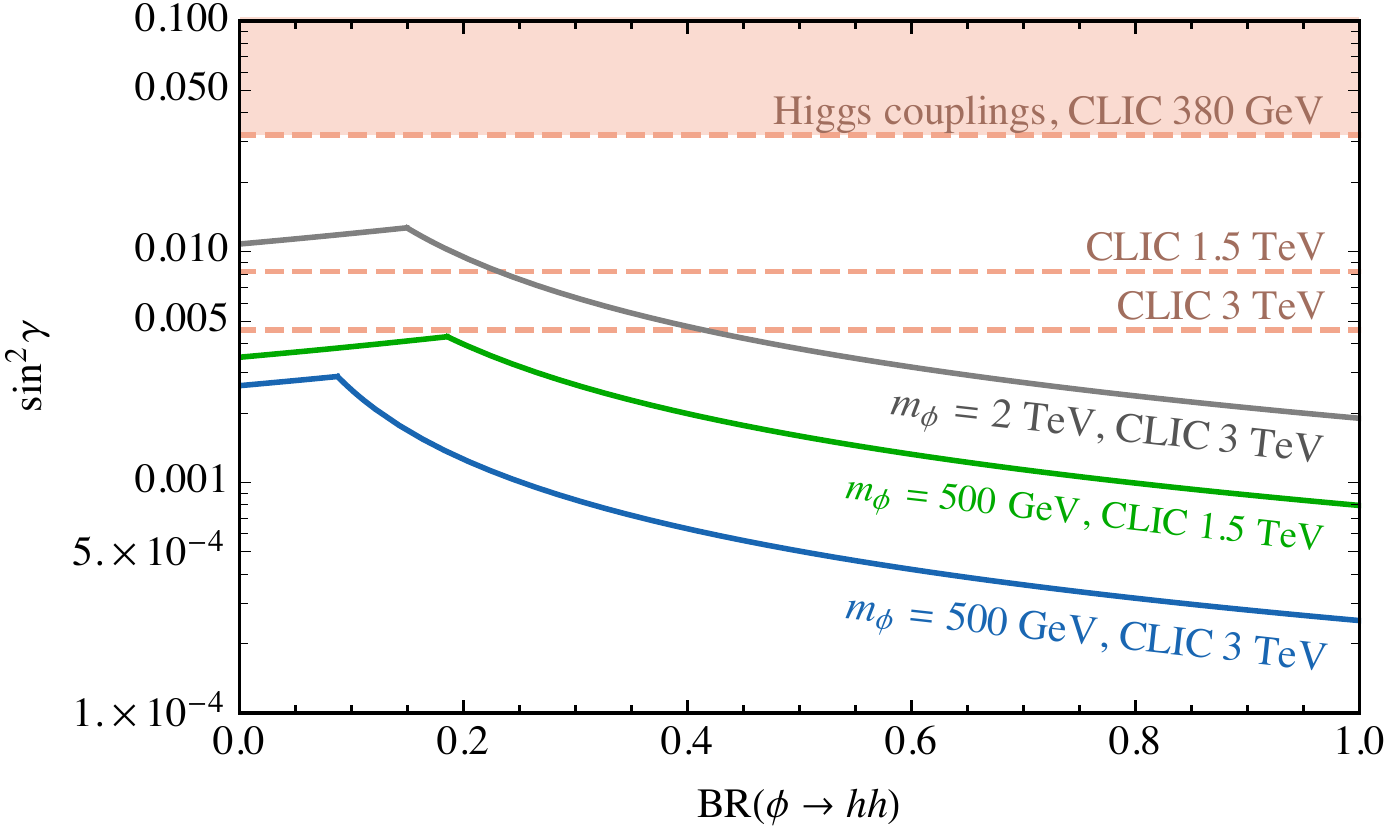}
\caption{Dependence on ${\rm BR}_{\phi\to hh}$ of the combined direct reach in $hh$ or $VV$ at CLIC, compared with the sensitivity to $\sin\gamma$ expected from Higgs couplings measurements taken from Ref.~\cite{CLIC:2016zwp}. The green and blue lines are for $m_\phi = 500$ GeV at CLIC Stage II and III, respectively, the gray dashed line is for $m_\phi = 2$ TeV (at 3 TeV). The pink lines show the precision in Higgs couplings at the three CLIC stages.\label{fig:singammaBR}}
\end{figure}

For each signal mass $m_\phi$ and final state $f$, we select the simulated background events with an invariant mass of the two vectors that falls within a window of width $\Delta_f$ around the resonance peak, $m_\phi -\Delta_{f} < m_{VV} < m_\phi + \Delta_{f}$. We take $\Delta_{4\ell} = 5\%$, $\Delta_{2\ell2j} = 10\%$, and $\Delta_{4j}=15\%$ for the three channels under consideration. The first two numbers are in rough agreement with the resolutions that are achieved at the LHC for resonances reconstructed in $ZZ$~\cite{Aad:2015kna}, while the third number correctly reproduces our results for $hh(4b)$ of the previous section.
The expected sensitivities are then computed from the resulting number of background events, solving Eq.~\eqref{eq:significance} to find the excluded number of signal events at 95\% C.L. (one sided), rescaled to take into account the branching ratios of $W$ and $Z$ into the relevant final states.
We show the sensitivities obtained this way in Figure~\ref{CLIClimits}, for the exclusive channels $ZZ(4\ell)$ and $ZZ(2\ell 2j)$, and for the combination of the channels $ZZ(4\ell)$, $ZZ(2\ell 2j)$, $ZZ(4j)$, and $WW(4j)$.

\subsection{Discussion and comparison with hadron colliders}\label{sec:hadron}

We now translate the projected sensitivities on the cross-section into a limit on the mixing angle $\sin^2\gamma$, which is trivially obtained rescaling by the cross-section of a SM Higgs with the same mass times the singlet branching ratio into $hh$ or $VV$.
For CLIC we compute the Higgs production cross-section using {\sc Madgraph 5} at LO, see Figure~\ref{fig:rate-theory} left.

In Figure~\ref{fig:singammaBR} we show the combined reach from the $hh(4b)$ and $VV$ searches described in the previous section, as a function of ${\rm BR}_{\phi\to hh}$ and compare it with the expected reach in Higgs signal strengths at the various stages of CLIC. Notice that, especially for lower singlet masses, direct searches are more powerful than indirect ones at each stage. Varying the value of the ${\rm BR}_{\phi\to hh}$, the searches in the two channels $hh$ and $VV$ are clearly complementary.

\begin{figure}[t!]
\centering
\includegraphics[width=0.7\textwidth]{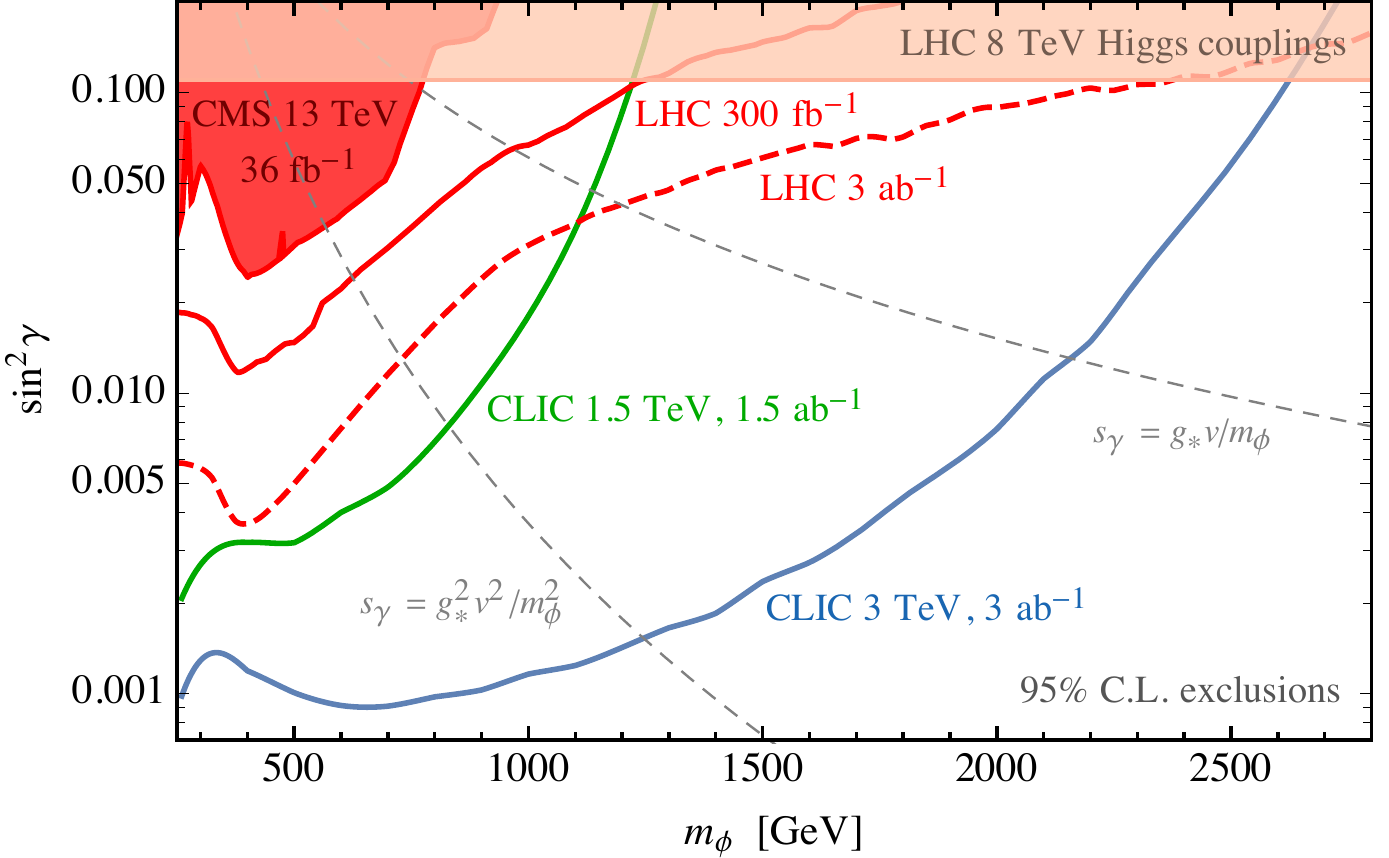}\hfill%
\caption{\label{fig:comparison} Exclusions at 95\% C.L. in the plane $(m_\phi,\sin^2\gamma)$. The shaded regions are the present constraints from LHC direct searches for $\phi\to ZZ$ (red) and Higgs couplings measurements (pink). The reach at CLIC Stage II (green) and Stage III (blue) in $\phi \to hh (4b)$ is compared with the projections for LHC in $\phi \to ZZ$ with a luminosity of 300 fb$^{-1}$ (solid red) and 3 ab$^{-1}$ (dashed red). We have fixed ${\rm BR}_{\phi\to hh} = {\rm BR}_{\phi\to ZZ} = 25\%$. The dashed grey lines show two different scalings of $s_\gamma$ with $m_\phi$, as described in Section~\ref{sec:scaling} ($g_* = 1$ in both cases).}
\end{figure}

We then compare our CLIC results with Higgs couplings measurements and direct searches at various stages of the LHC: the current run at 13 TeV, the end of the 14 TeV run with 300 fb$^{-1}$, and the high-luminosity phase with 3 ab$^{-1}$.
For what concerns the Higgs couplings, we display as excluded the combined ATLAS and CMS 8 TeV constraint~\cite{Khachatryan:2016vau}, in order to be conservative.
Indeed, the 13 TeV best-fit Higgs signal strength from CMS~\cite{CMS-PAS-HIG-17-031} is larger than the SM one by almost two sigma.
For direct searches we show the present LHC exclusions~\cite{Sirunyan:2018qlb,Aaboud:2017itg} as well as the projected sensitivities at 300 fb$^{-1}$ and 3 ab$^{-1}$.

To determine future sensitivities at $pp$ colliders, we rescale the expected sensitivity of the existing 13~TeV search~\cite{Sirunyan:2018qlb} at higher energies and luminosities using quark parton luminosities, with a procedure analogous to the one presented in Ref.~\cite{Buttazzo:2015bka}.
We refer the reader to Appendix~\ref{app:lepton} for more details on this aspect, as well as for the expected sensitivities also at the HE-LHC with center-of-mass energy of 27 TeV, and FCC-hh at 100 TeV, that will be used in the next section when comparing to muon colliders.

To translate the future sensitivities on cross-section to a reach in mixing angle, we use the SM Higgs production cross-section at hadron colliders. We include gluon fusion plus VBF plus VH production, and compute the first one with {\sc ggHiggs v3.5}~\cite{Ball:2013bra,Bonvini:2014jma,Bonvini:2016frm}, and the second two with {\sc Madgraph 5} at LO.

It can be seen that CLIC at 1.5 TeV and 3 TeV will be able to probe singlet masses up to about 1.2 TeV and 2.6 TeV, respectively. CLIC at 1.5 TeV stage will be more sensitive than the high-luminosity LHC up to masses of about 1 TeV, while the 3 TeV stage will be significantly more sensitive over the full mass range.

\subsection{Muon colliders}\label{sec:muon_colliders}

Among the proposals for muon colliders, the two main categories differ in the way muon bunches are produced: either from protons on target (MAP~\cite{Delahaye:2013jla}), or from positrons on target (LEMMA~\cite{Antonelli:2013mmk,Antonelli:2015nla,Collamati:2017jww}).
The former case guarantees a larger instantaneous luminosity, but is complicated by the technological challenge of muon cooling. The latter case overcomes this difficulty at the price of a slightly reduced luminosity. See Ref.~\cite{padova} for an updated discussion about the state-of-art of muon colliders.

Here and in the following we simply focus on an idealised muon collider at 6~TeV and 14~TeV, respectively with 6~ab$^{-1}$ and 14~ab$^{-1}$ of integrated luminosity, as benchmarks attainable both by MAP and LEMMA. We determine their sensitivity to resonances decaying to $hh(4b)$ only at the {\sc Madgraph} level, because of the present lack of knowledge of the detectors that will be used at those machines.

\begin{figure}[t]
\centering
\includegraphics[width=0.7\textwidth]{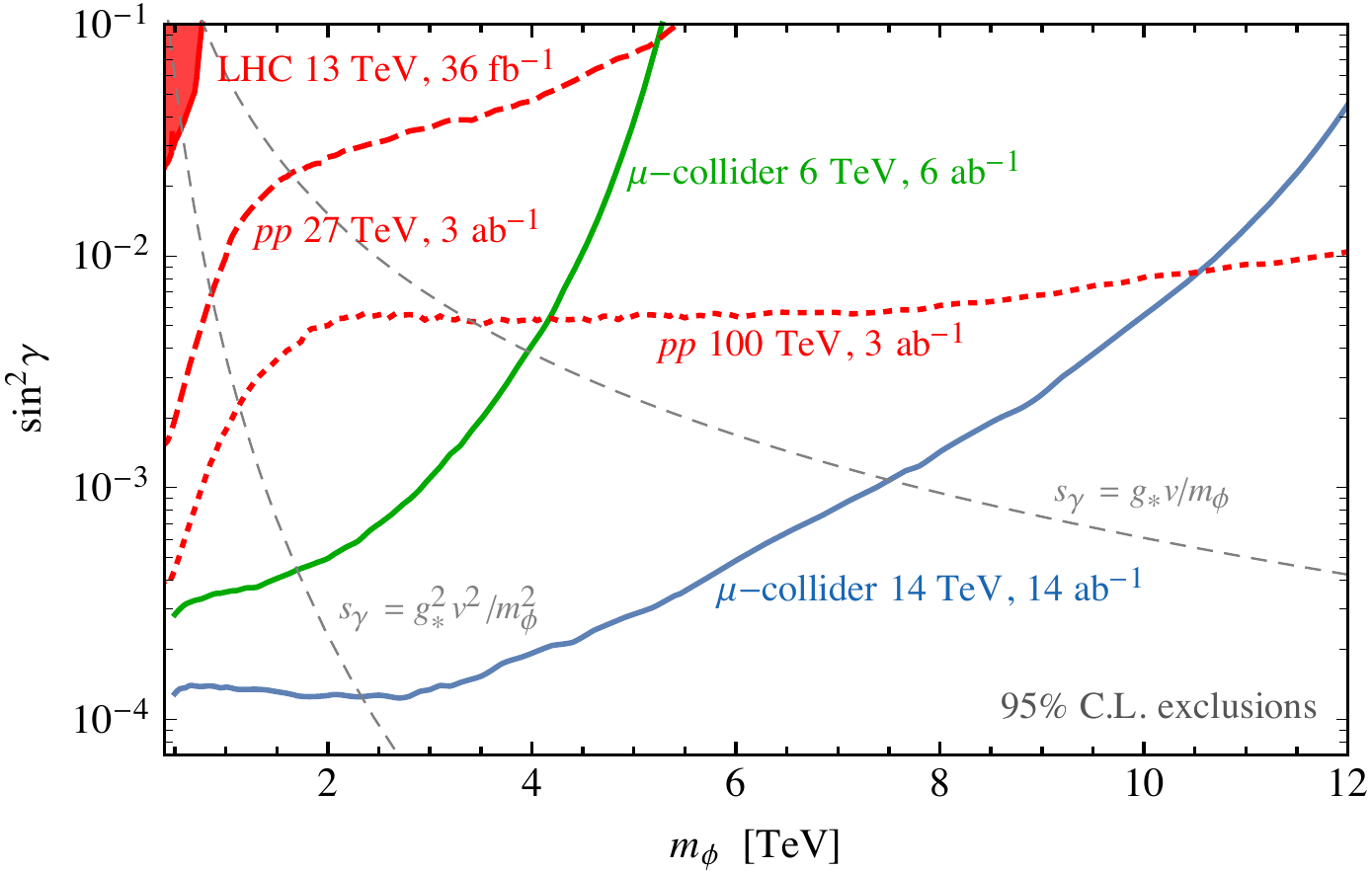} 
\caption{Sensitivies of very high energy hadron and muon colliders at 95\% C.L. in the plane $(m_\phi,\sin^2\gamma)$.
The red lines show the reach in $\phi \to ZZ$ of HE-LHC at $\sqrt{s}=27$ TeV (dashed) and FCC-hh at $\sqrt{s}=100$ TeV (dotted), both with 3 ab$^{-1}$.
The solid lines show the reach in $\phi \to hh (4b)$ of a muon collider at $\sqrt{s}=6$ TeV with 6 ab$^{-1}$ (green), and at $\sqrt{s}=14$ TeV with 14 ab$^{-1}$ (blue).
We have fixed ${\rm BR}_{\phi\to hh} = {\rm BR}_{\phi\to ZZ} = 25\%$. The grey dashed lines show two possible scalings for $s_\gamma$, as described in Section~\ref{sec:scaling} ($g_* = 1$ in both cases). \label{futuro}}
\end{figure}

We simulate the background processes $\mu^+\mu^- \to \nu \bar{\nu} hh$ and impose the cut $|\eta_h| < 2$ on the pseudo-rapidity of each Higgs boson, which roughly corresponds to $|\cos\theta_h| < 0.95$. For every signal mass $m_\phi$ that we want to test, we then take the fraction of background events that satisfies $m_{hh} = m_\phi \pm 15\%$.
Finally we assume an additional efficiency of 30\% (as a rough estimate of $b$ identification and other effects), and we determine the 95\%C.L.\ sensitivities according to Eq.~\eqref{eq:significance}, using as usual systematics of 2\%, and of course taking into account the branching ratio of $h \to b\bar{b}$.
For the signals, we compute the total $\ell^+\ell^- \to \nu \bar{\nu} \phi$ cross-sections at LO with {\sc Madgraph}, at all the machines of our interest.
We then just impose the same efficiency of 30\%, assuming it captures the effects of the various cuts (this assumption is to some extent supported by the study in Section~\ref{sec:Stohh}).

We find that this procedure reproduces extremely well, at both 1.5 and 3 TeV, the results of the more careful detector study of Section~\ref{sec:Stohh}, at least for $m_{hh} \gtrsim 700$~GeV.
We report in Appendix~\ref{app:lepton} more details on the validation above, as well as the sensitivities both on the mixing angle and on the production cross-section of a generic resonance decaying to $hh$, at lepton machines from 1.5 TeV to 14 TeV of center-of-mass energy.
Since these searches are essentially background-free for large masses, they are dominated by statistical errors. We discuss the impact of systematic errors in more detail in Appendix~\ref{app:lepton}, also in relation with possible target luminosities for muon colliders.

Here, we show~in Figure~\ref{futuro} the 95\% C.L.\ sensitivities in the plane $(m_\phi,\, \sin^2\gamma)$ at $\sqrt{s}=6$ TeV and 14 TeV, for total integrated luminosities of 6 ab$^{-1}$ and 14 ab$^{-1}$, respectively. We also compare the reach of muon colliders to the one of high-energy hadron collider proposals such as HE-LHC and FCC-hh.
The take-home message of this comparison is that HELCs in the very high energy regime could become very powerful discovery machines, even stronger than future hadronic colliders, at least for New Physics mostly coupled to the Higgs sector.

\section{Single Production \& Beyond the Standard Model Scenarios}\label{sec:models}

In this section we discuss the implication of the CLIC reach on singlet resonances in well motivated Beyond the Standard Model (BSM) scenarios. 

\subsection{NMSSM}\label{sec:NMSSM}

In the NMSSM, the particle content of the MSSM is extended with a singlet of the SM gauge group $S$, so that the superpotential reads $W = W_{\mathsmaller{\rm MSSM}} + \lambda\,S H_u H_d + f(S)$, with $f$ a polynomial up to degree~3.
The SM-like Higgs boson mass receives an extra tree-level contribution, which lifts its upper limit to
\be
m_h^2 < m_Z^2 \,\cos^2 2\beta + \lambda^2 \,v^2\,\sin^2 2\beta/2 + \Delta_{hh}^2\,,
\ee
where $t_\beta = \tan\beta$ is the ratio between the up and down Higgs VEVs, and $\Delta_{hh}$ encodes the usual SUSY radiative contributions.
In light of the null LHC searches for coloured superpartners, the NMSSM with a large coupling $\lambda$ is a particularly attractive SUSY model from the point of view of naturalness of the EW scale, see e.g.\ Ref.~\cite{Barbieri:2006bg,Hall:2011aa,Agashe:2012zq,Gherghetta:2012gb}.
Indeed, the fine-tuning needed to reproduce the EW scale is parametrically alleviated, for a fixed value of the stop and gluino masses and with respect to the MSSM, by a factor $\sim \lambda/g$.
Having $\lambda \gtrsim 1.5$ would however overshoot the Higgs mass, and thus introduces a new tuning problem to bring $m_h$ down to its measured value.
In addition, $\lambda \simeq 2$ becomes strongly coupled at scales of $\mathcal{O}(10)$ TeV.\footnote{The additional requirement that $\lambda$ be perturbative up to the GUT scale imposes, at the weak scale, $\lambda \lesssim 0.7$~\cite{Espinosa:1991gr}. However, it is conceivable that a strong sector exists at the scale where $\lambda$ becomes non-perturbative, and without affecting the success of GUT in the NMSSM, see e.g.\ the model in Ref.~\cite{Barbieri:2013hxa} and references therein.}

\begin{figure}[t!]
\centering
\includegraphics[width=0.49\textwidth]{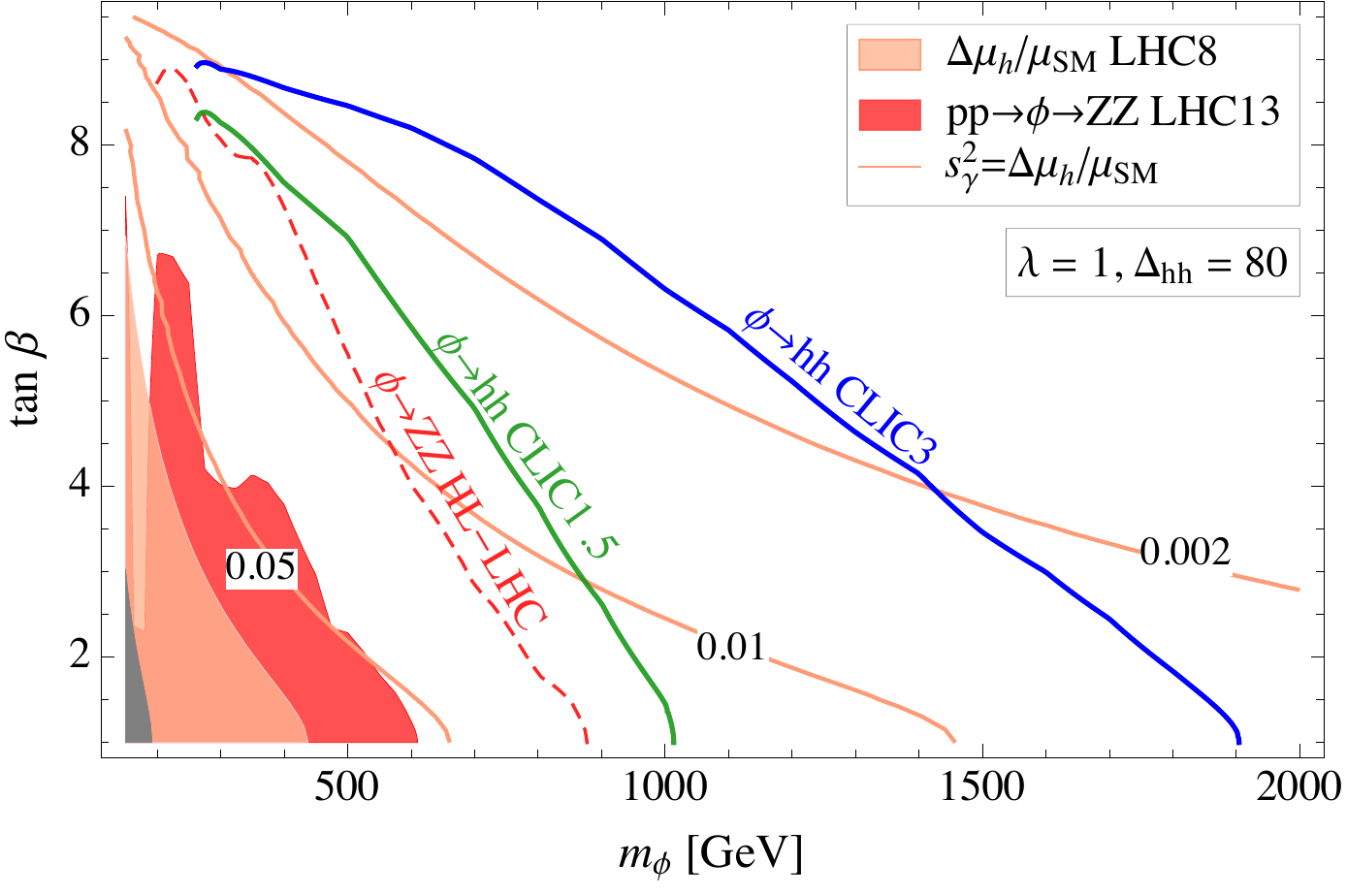}\hfill%
\includegraphics[width=0.5\textwidth]{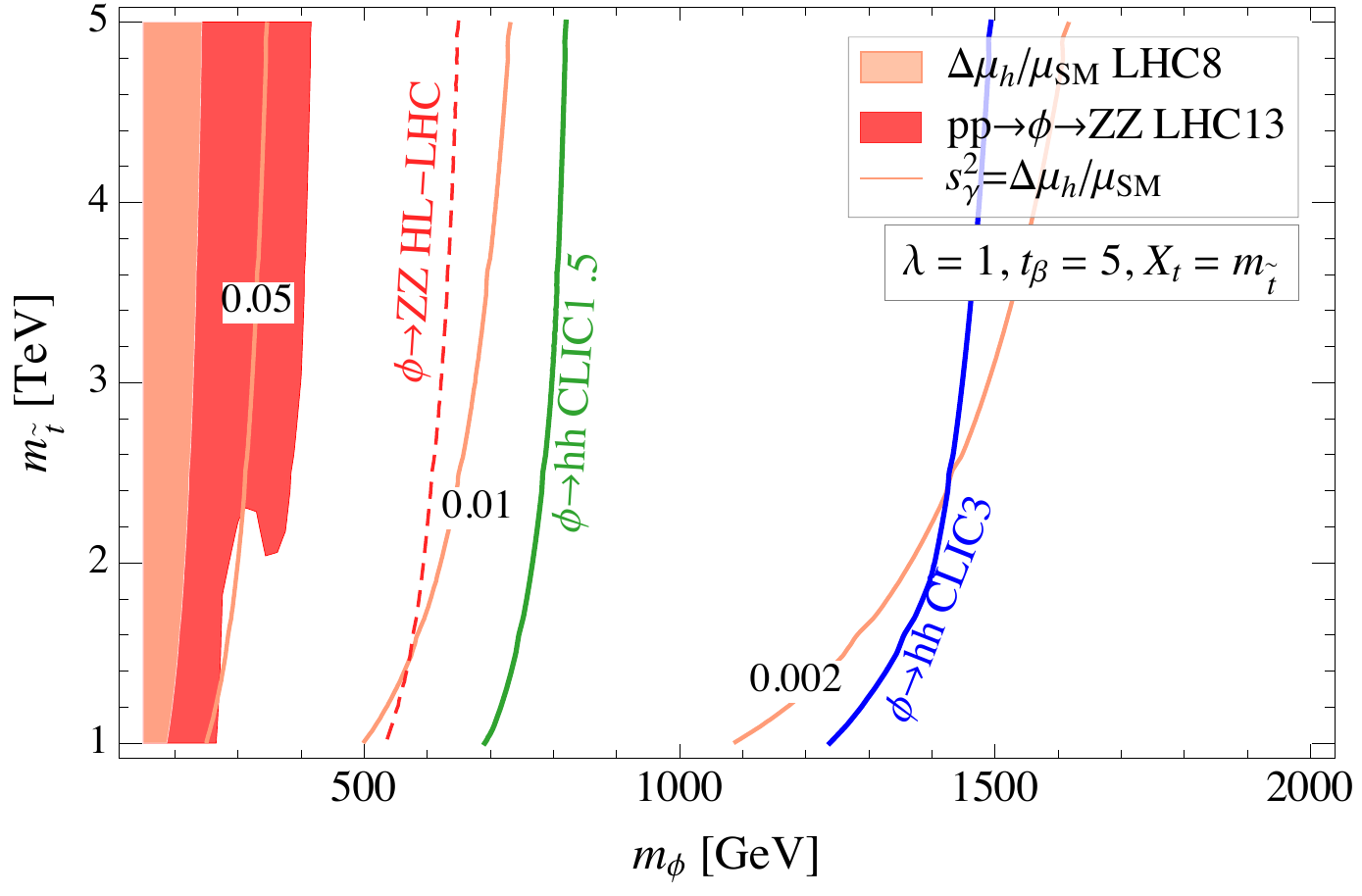} 
\caption{\label{fig:NMSSM} Constraints at 95\% C.L. The shaded regions are the present constraints from LHC direct searches for $\phi\to ZZ$ (red) and Higgs couplings measurements (pink). The reach at CLIC Stage 2 and Stage 3 (blue) is compared with the projections for LHC with a luminosity of 3 ab$^{-1}$ (dashed red). We have fixed ${\rm BR}_{\phi\to hh} = {\rm BR}_{\phi\to ZZ} = 25\%$. {\bf Left:}  Plane $(m_\phi, \tan \beta)$ {\bf Right:} Plane $(m_\phi, m_{\tilde{t}} = \sqrt{m_{\tilde{t}_1}\,m_{\tilde{t}_2}})$. }
\end{figure}

The NMSSM constitutes an ideal application of our sensitivity study for extra singlet scalars.
To describe the scalar sector of the NMSSM we employ the parametrisation put forward in Ref.~\cite{Barbieri:2013hxa,Barbieri:2013nka}.
Then, to obtain a simple description of the scalar singlet phenomenology, we assume the extra Higgs doublet to be in the decoupled or alignment limit. This is also somehow favoured by the present LHC constraints, because the mixing of the SM Higgs with a doublet is more constrained than the mixing with a singlet, see e.g.\ Ref.~\cite{Barbieri:2013hxa}. The phenomenology of the SM-like Higgs plus the extra singlet can then be described by 4 free parameters
\be
m_\phi, \quad t_\beta, \quad \lambda,\quad \Delta_{hh}.
\label{eq:singlets_NMSSMpars}
\ee
To connect $\Delta_{hh}$ with parameters of more immediate physical interpretation, we employ the concise analytical expression (see e.g.\ Ref.~\cite{Blum:2012ii,Drees:2004jm})
\be
\Delta_{hh} = \frac{3}{2 \pi^2}\,\frac{1}{v^2}\,\left( m_t^4(Q_t) \log\frac{m_{\tilde t}^2}{M_t^2} + m_t^4(m_{\tilde t}) \, \frac{X_t^2}{m_{\tilde t}^2}\,\Big( 1- \frac{1}{12}\frac{X_t^2}{m_{\tilde t}^2}\Big)\right),
\ee
where $m_{\tilde{t}} = \sqrt{m_{\tilde{t}_1}\,m_{\tilde{t}_2}}$, $m_{\tilde{t}_{1,2}}$ are the physical stop masses, $X_t = A_t -\mu/t_\beta$,  $M_t = 173$~GeV is the top pole mass, $m_t(Q)$ is the running top mass, and $Q_t = \sqrt{M_t m_{\tilde{t}}}$. Such expression is accurate to the level of a few GeV in $\Delta_{hh}$, which is more than enough for our purposes.

The phenomenology of the Higgs plus singlet system is displayed, for the NMSSM, in Figure~\ref{fig:NMSSM}, where we fix $\lambda =1$ as a benchmark motivated by naturalness.
In the left-hand panel we let $m_\phi$ and  $t_\beta$ vary and we fix $\Delta_{hh} = 80$~GeV, a value obtainable e.g. for stop masses in the range of 1-2 TeV. The precise value of $\Delta_{hh}$ does not affect the Higgs sector phenomenology as long as it is within O(10\%) of 80 GeV.
One sees that direct searches for the extra singlet, at both CLIC stages II and III, would probe a parameter space that is completely unexplored by the HL-LHC. At CLIC, direct $\phi$ searches and Higgs coupling measurements would constitute a complementary probe of the parameter space, as noticed already in Section~\ref{sec:single}. 

To connect with the phenomenology of the SUSY coloured sector, in the right-hand panel we let $m_\phi$ and $m_{\tilde t}$ vary, for $t_\beta = 5$ and $X_t = m_{\tilde t}$. We see that the precise value of the stop masses does not have a major impact on the phenomenology of the Higgs-singlet sector. This consideration holds independently of the values $t_\beta$ and $X_t$, with the only exception of large $t_\beta$ and very small $X_t$ (in which one recovers the MSSM problems in reproducing the correct Higgs mass). We can therefore conclude that searches for the singlet scalar will not provide significant information about the stop masses. They will, on the other hand, give a measure of the tuning in the scalar sector, which is a dominant source of tuning in the NMSSM. The contribution due to the singlet can be roughly quantified as $\lambda^2 s^2/m_Z^2 \propto m_\phi^2/m_Z^2$ (where $s$ is the VEV of the singlet).\footnote{The tuning due to the second doublet also grows with its mass, and is an independent contribution.}

Here we have not discussed deviations in the trilinear Higgs coupling in the NMSSM. They depend on more parameters than those in Eq.~\eqref{eq:singlets_NMSSMpars}, and can reach O(50\%) or more if $\lambda \gtrsim 1$, see Ref.~\cite{Sala:2015lza} for a precise quantification.

\subsection{Twin Higgs}\label{sec:TH}

In Twin Higgs models the SM Higgs sector is extended by adding the twin Higgs $H_B$, which is a singlet under the SM and a doublet under a mirror EW gauge group $SU(2)_B$. The twin Higgs is coupled to the SM Higgs $H_A$ via a portal coupling $\lambda_*$, that realises a global SO(8) symmetry at tree-level. This is spontaneously broken to SO(7) at a scale $f$.
The radiative stability of the construction is ensured by an approximate $Z_2$ symmetry between the SM and the mirror sector, so that the full content of the SM, or part of it, is doubled~\cite{Craig:2015pha}. 
An explicit breaking of the $Z_2$ symmetry is then introduced to allow for $f > v$ and for a viable phenomenology~\cite{Chacko:2005pe}.
All in all, the scalar potential reads
\be
V= \lambda_*\Big(|H_A|^2+|H_B|^2 -\frac{f_0^2}{2}\Big)^2 +  \kappa (|H_A|^4+|H_B|^4) + \sigma_{\rm soft} f^2\, |H_A|^2 + \rho_{\rm hard} |H_A|^4\,,
\label{eq:V_TH}
\ee
where $\kappa$ is the SO(8)-breaking quartic and we parametrise the $Z_2$-breaking contributions by $\sigma_{\rm soft}$ and $\rho_{\rm hard}$. These correspond to a \emph{soft} and a \emph{hard} breaking of the discrete symmetry respectively. 
Notice that $\kappa$ receives irreducible IR contributions from (mirror) top-loops, and that if $Z_2$ is broken only softly at some scale, then a small quartic $\rho_{\rm hard}$ is generated at lower energies by $Z_2$-breaking 1-loop thresholds.

\begin{figure}[t!]
\centering
\includegraphics[width=0.5\textwidth]{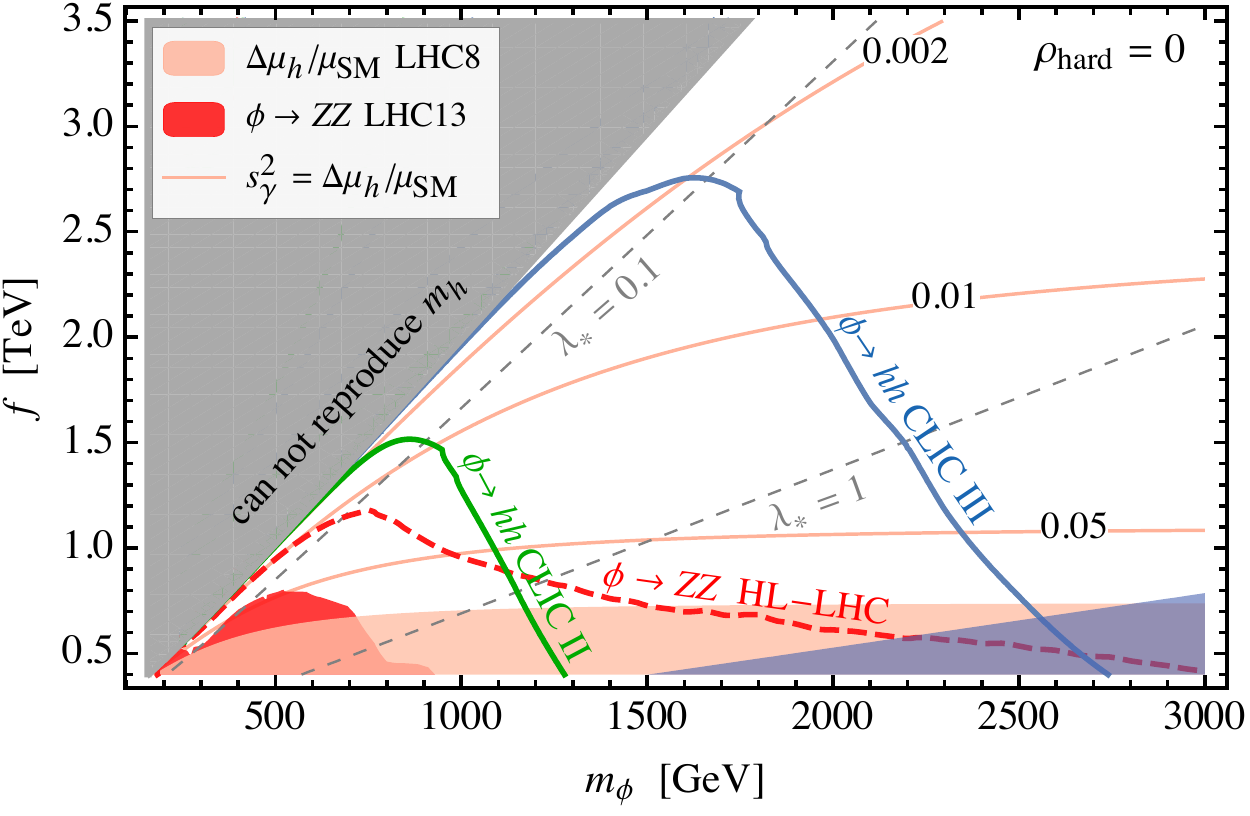}\hfill%
\includegraphics[width=0.49\textwidth]{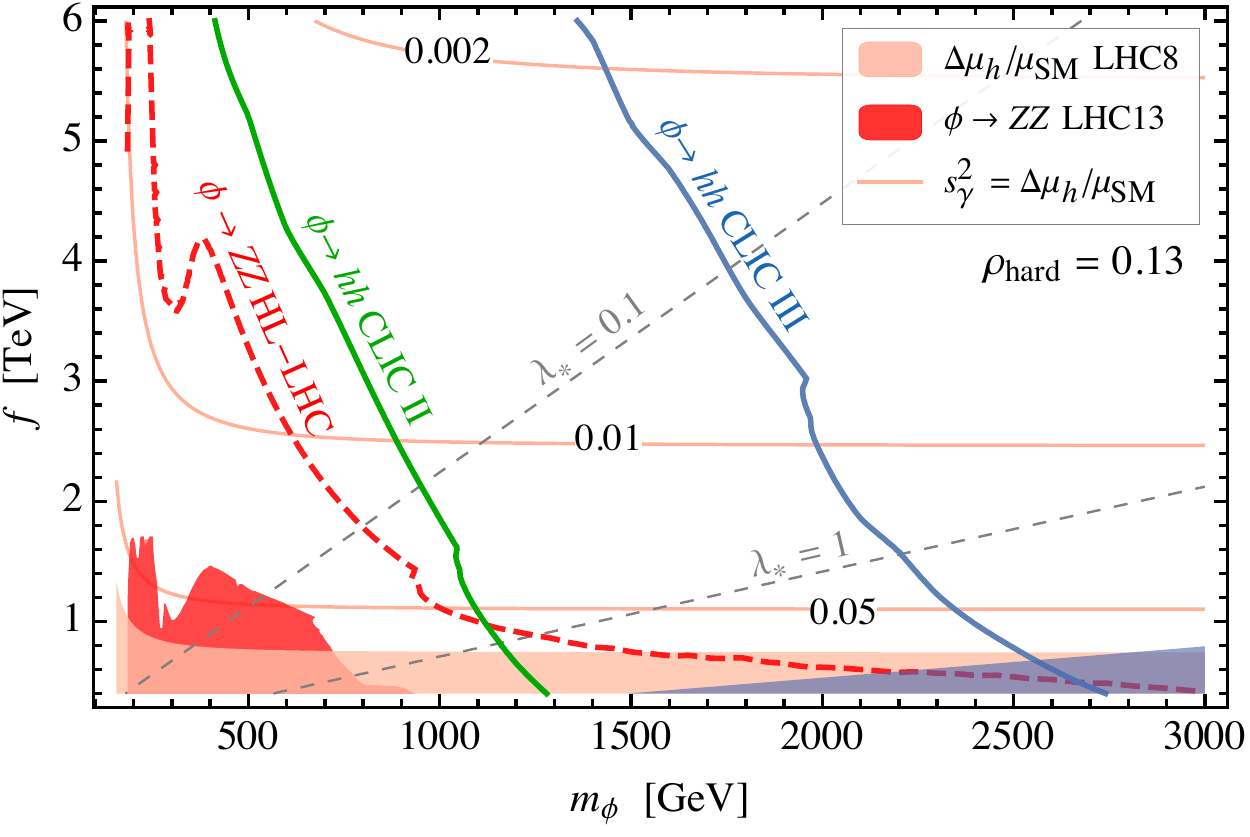} 
\caption{\label{fig:TH} Constraints at 95\% C.L. in the plane $(m_\phi, f)$. The shaded regions are the present constraints from LHC direct searches for $\phi\to ZZ$ (red) and Higgs couplings measurements (pink), and the region where $\Gamma_\phi > m_\phi$ (dark blue). The reach at CLIC Stage 2 and Stage 3 (blue) is compared with the projections for LHC with a luminosity of 3~ab$^{-1}$ (dashed red).The dashed grey lines correspond to constant values of the SO(8)-symmetric quartic $\lambda_*$. {\bf Left:} Only soft $Z_2$ breaking $\rho_{\rm hard} = 0$. {\bf Right:} A hard $Z_2$ breaking quartic $\rho_{\rm hard} = 0.13$.}
\end{figure}

After the spontaneous breaking of the SO(8) symmetry, we are left with two real scalars in the spectrum: the SM-like Higgs $h$ and the radial mode~$\phi$. Their physical masses, in the limit $\lambda_\ast\gg \kappa\, ,\sigma_{\rm soft}\, ,\rho_{\rm hard}$, read
\begin{align}
&m_\phi^2 \simeq 2 \lambda_* f^2\ ,\qquad\qquad m_h^2\simeq 2 v^2(2\kappa+\rho_{\rm hard})\ .
\end{align}
This explicitly shows that the Higgs mass is of the correct size for typical values of the parameters $\kappa$ and $\rho_{\rm hard}$.
In TH models the fine-tuning is parametrically reduced with respect to the ones of regular SUSY or Composite Higgs scenarios by $\lambda_H/\lambda_\ast$, where $\lambda_H\simeq 0.13$, see e.g.\ Ref.~\cite{Katz:2016wtw,Contino:2017moj}.
In models where the $Z_2$-breaking is mostly achieved by the quartic $\rho_{\rm hard}$,
one obtains an additional gain in fine-tuning by $\lambda_H/\vert\lambda_H-\rho_{\rm hard}\vert$, which is maximised for $\rho_{\rm hard}$ as close as possible to the SM quartic $\lambda_H$.
This parametric gain is however limited by the irreducible IR contributions to the SO(8)-breaking quartic $\kappa$, as discussed in Ref.~\cite{Katz:2016wtw}.

The requirement to reproduce the EW scale $v$ and the Higgs mass $m_h$ fixes 2 out of the 5 free parameters in Eq.~\eqref{eq:V_TH}.
We choose the three remaining free parameters as the spontaneous breaking scale $f$, the physical singlet mass $m_\phi$, and the $Z_2$-breaking quartic $\rho_{\rm hard}$.
We then find the following exact analytical expression for the mixing angle
\begin{equation}
\sin^2\gamma =  \frac{v^2}{f^2} -\frac{m_h^2 }{m_\phi^2-m_h^2}\Big(1-2\frac{v^2}{f^2}\Big) + \frac{2 \rho_{\rm hard} v^2}{m_\phi^2-m_h^2} \Big(1-\frac{v^2}{f^2}\Big)\,,
\label{eq:sinsqTH}
\end{equation}
which to our knowledge was never presented before in the literature.
Other useful relations, to track the impact of $\rho_{\rm hard}$, are
\begin{equation}
\dfrac{g_{hhh}}{g_{hhh}^{\rm SM}} \simeq 1 - \frac{3}{2} \frac{v^2}{f^2} \left[1 -\frac{\lambda_H - \rho_{\rm hard}}{\lambda_*} \right], \qquad\qquad\qquad g_{\phi hh} \simeq \frac{m_\phi^2}{f} \left[1- \frac{\lambda_H - \rho_{\rm hard} }{2 \lambda_*} \right], \label{eq:gphihh}
\end{equation}
where $g_{hhh}^{\rm SM} = 3m_h^2/2v$, and where we ignored all higher orders in $v^2/f^2$ and $\lambda_H/\lambda_* \approx \rho_{\rm hard}/\lambda_*$.
Eqs~(\ref{eq:sinsqTH})--(\ref{eq:gphihh}) show nicely that the effect of the new quartic decouples with the mass of the singlet state (or equivalently with $\lambda_*$), and therefore it could affect the phenomenology of the scalar sector only at small to intermediate $m_\phi$. 

The parameter space of TH models is displayed in Figure~\ref{fig:TH} in the $m_\phi - f$ plane, for two benchmark values of the hard breaking quartic $\rho_{\rm hard}$.\footnote{For previous related phenomenological studies of the radial mode in TH see Ref.~\cite{Buttazzo:2015bka,Ahmed:2017psb}.}
As anticipated by the analytical understanding above, the region where the impact of $\rho_{\rm hard} \neq 0$ is most visible is the one where $m_\phi$ is small. In particular we see that a non-zero $\rho_{\rm hard}$ allows the Higgs mass constraint to be satisfied at large $f$ and small $m_\phi$. In this region the Higgs mass is mostly achieved via $\rho_{\rm hard}$. However, in the same region the fine tuning gain of the TH is limited because $\lambda_* \lesssim 0.1$~\cite{Katz:2016wtw}.

Figure~\ref{fig:TH} also displays the phenomenological results of Section~\ref{sec:single}, where we have extended the framework to include the invisible decays of the radial mode into $W'W'$, $Z'Z'$ (all with masses $m_W \times f/v$, because the $U(1)'$ could well be not gauged~\cite{Barbieri:2015lqa,Low:2015nqa}) and $t'\bar{t}'$ (with mass $m_t \times f/v$). The $SO(8)$ symmetry implies that the invisible branching ratio asymptotises to 3/7 for $m_\phi \gg m_t'$.
One learns from Figure~\ref{fig:TH} that the phenomenology of the twin Higgs $\phi$ is independent on how the $Z_2$-breaking is achieved, at least in the region of parameter space where the fine-tuning is ameliorated. HELCs like CLIC are expected to probe the most natural regions of TH models mainly via their precision in Higgs coupling measurements. While direct searches for the radial mode would constitute a weaker probe of the interesting region of the parameter space, they could provide precious complementary information. A similar conclusion was drawn also in \cite{Chacko:2017xpd}, where the $hh(4b)$ signature was studied.

\subsection{Comments on heavy electroweak ALPs}\label{sec:ALP}

\begin{figure}[t]
\begin{center}
\includegraphics[width=.50\textwidth]{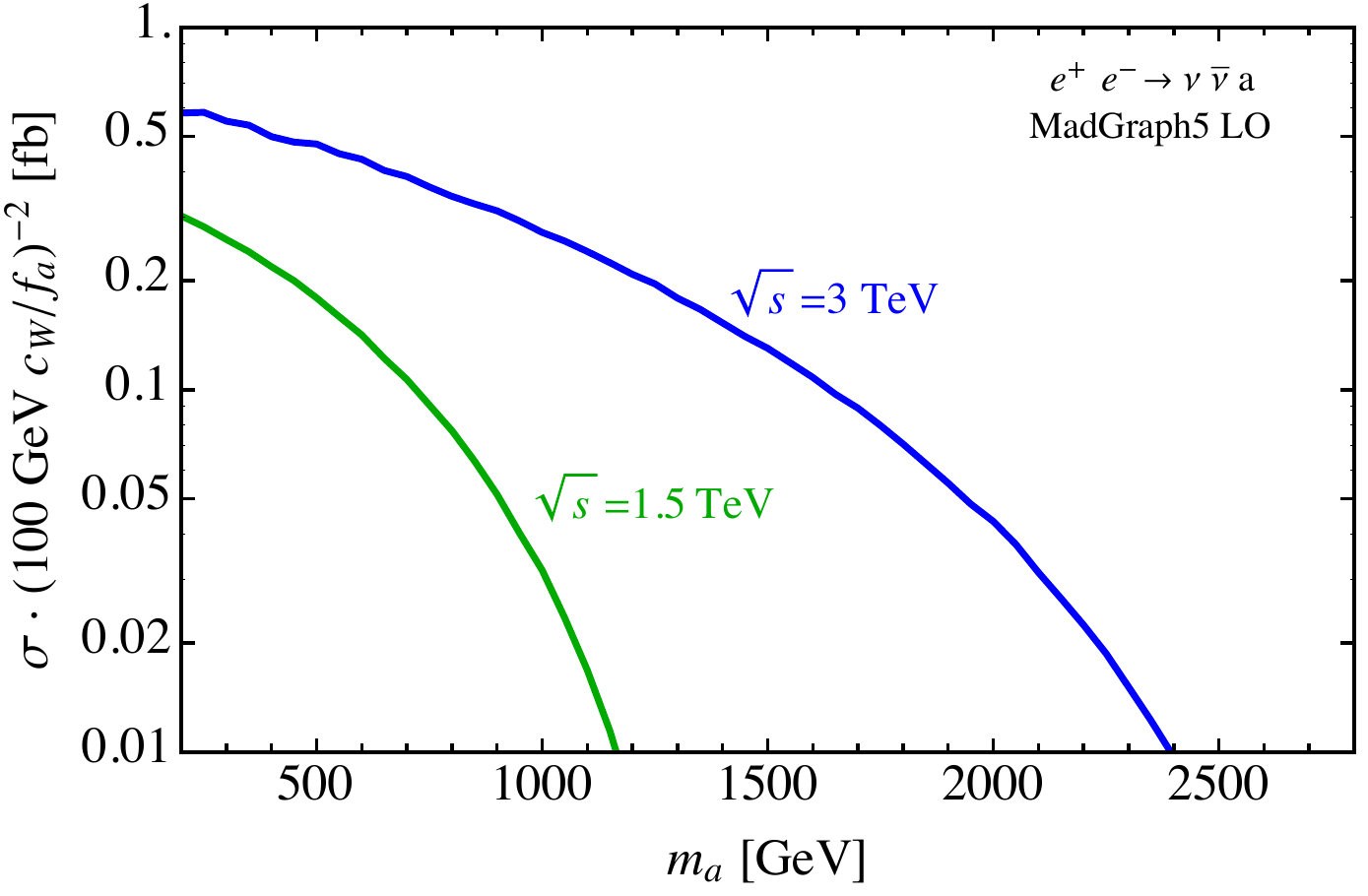}\hfill%
\includegraphics[width=.48\textwidth]{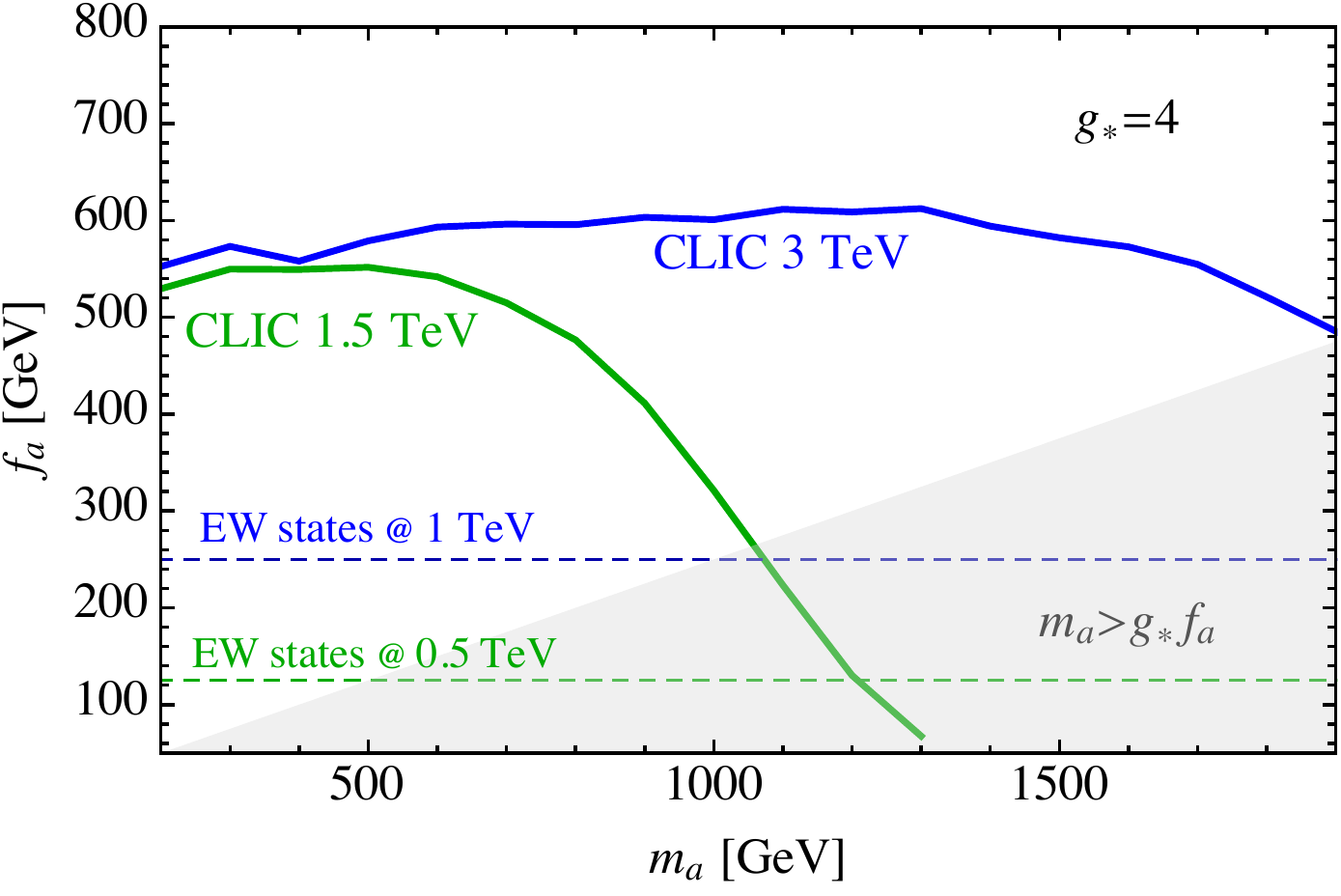} 
\caption{{\bf Left:} single production via $WW$-fusion of a photophobic ALP $c_1=-3/5 c_2$. {\bf Right:} Reach of CLIC at 1.5 TeV (green ) and 3 TeV (blue) in the photophobic ALP parameter space for $g_*=4$ $(c_2=16\pi^2/g_\ast^2)$. In grey we show the region where $m_a\gtrsim g_* f$ and the EFT in Eq.~\eqref{eq:ALPeft} ceases to be justified. Dashed lines indicate the scale of the EW states which could be within the reach of CLIC at 1.5~TeV (green) and at 3~TeV (blue). \label{fig:ALP}}
\end{center}
\end{figure}

Our results can be applied generically also to scalar resonances that are produced singly from the fusion of transverse $W$ bosons. Resonances of this type are the so-called axion-like particles (ALPs), a quite generic category of pseudo-scalar particles coupled via ABJ anomalies to the SM gauge bosons. These arises in many theoretical models related to Dark Matter production \cite{Nomura:2008ru}, Naturalness \cite{Ferretti:2013kya,Gripaios:2009pe,Bellazzini:2017neg} and vector-like confinement \cite{Kilic:2009mi}.

In this context we consider a somehow heavy ALP $a$ with only electroweak anomalies and mass $m_a> 2 m_W$. The effective Lagrangian for an ALP of this type reads  
\begin{align}
\mathscr{L}_{\text{ALP}}&=\frac{1}{2}(\partial_\mu a)^2 -\frac{1}{2}m_a^2 a^2 + \frac{c_1\alpha_1}{4\pi}\frac{a}{f_a} B\tilde{B}+\frac{c_2\alpha_2}{4\pi}\frac{a}{f_a} W\tilde{W}\,, \label{eq:ALPeft}
\end{align}
where $\tilde{F}^{\mu\nu}=(1/2)\,\epsilon^{\mu\nu\rho\sigma}F_{\rho\sigma}$ for any field strength, $\alpha_1=5\alpha_Y/3$ is the GUT-normalised $U(1)_Y$ coupling constant.  The scale $f_a$ is associated to the spontaneous breaking of the $U(1)$ under which the ALP shifts, while we do not specify the origin of the explicit breaking terms $m_a$.
The above theory has a physical cut-off at a scale $g_* f_a$, where both the radial mode and new particles charged under the electroweak group appear, so that for consistency $m_a < g_* f_a$.
The anomaly coefficients $c_{1,2}$ depend linearly on the number of states in the UV physics which carry EW quantum numbers and are charged under the global $U(1)$. After electroweak symmetry breaking (EWSB), one can write 
\begin{equation}
 c_{WW}=c_2, \quad c_{ZZ}=c_2+\tan\theta_W^4\frac{5}{3} c_1, \quad c_{Z\gamma}=c_2-\tan\theta_W^2\frac{5}{3} c_1,\quad c_{\gamma\gamma}=c_2+\frac{5}{3}c_1\,.
\end{equation}
Independently of the underlying UV theory, the anomaly coefficients are correlated with $g_*$.
For example, in a QCD-like sector with $N$ colours $g_*\sim 4\pi/\sqrt{N}$, and for a `composite' ALP from that sector one roughly expects $c_{1,2}\sim N$ from the degrees of freedom above the confinement scale.
Similarly, in weakly coupled models with $N_f$ charged fermions of masses $\sim g_*f_a$ one has $c_{1,2}\sim N_{f}$ (see e.g. Ref.~\cite{Bellazzini:2017neg}), and $g_* \lesssim 4\pi/\sqrt{N}$ from perturbativity.
The arguments above indicate that a large ALP coupling to gauge bosons requires a somehow small $g_*$, and therefore it strengthens the bound $m_a < g_* f_a$.

In order to focus on a concrete model, we choose as a benchmark the photophobic ALP discussed in Ref.~\cite{Craig:2018kne}, defined by $c_{\gamma\gamma}=0$. Notice that assuming a zero photon coupling is a good approximation even after radiative corrections are included. Analogous results can be obtained in more general models as long as $\vert c_2/c_1\vert$ is large enough to not make the branching ratios in vector bosons subdominant compared to the ones in final states containing photons. In the photophobic ALP the WW branching ratio dominates over the ZZ with $\text{BR}_{WW}/\text{BR}_{ZZ}\simeq 4$ at high masses. 

For this reason in the right panel of Figure~\ref{fig:ALP}  we show the reach in the plane $(m_a,f_a)$ of the $WW$ channel at the stage II and III of CLIC as determined by the analysis in Section~\ref{sec:StoVV}. The only difference compared to the cases discussed in previous sections is that the ALP couples only to the transverse polarisations of the gauge bosons. The cross sections for single production at 1.5 TeV and 3 TeV are given in the left panel of Fig.~\ref{fig:ALP}. These are computed for the signal $e^+e^- \to a(2V) \nu\bar\nu$ with {\sc MadGraph}, after implementing the Lagrangian in Eq.~\eqref{modello} in {\sc FeynRules} 2.0.We refer to~\cite{Dawson:1984gx} for a derivation of analogous formulas to the ones presented in Eq.~\eqref{single} for the transverse polarisations. 

The final reach depends largely on $c_2=16\pi^2/g_*^2$, which affects the rate at quadratic level, and it has been set to reproduce a benchmark with $g_*=4$.  A relatively small $g_*$ is required to have a sizeable production rate. However, in the same regime, the EW charged UV states responsible for generating the anomaly are pushed down to be within the reach of high-energy colliders.  Aware of this issue, we show the line corresponding to masses of new electroweak states of 0.5 and 1 TeV, which roughly indicates the reach of CLIC on the pair production of these particles at stage II and III respectively  (see e.g.\ Ref.~\cite{Baer:2013faa}). Notice that this is somehow less transparent in the parametrisations of other studies (e.g.\ Ref.~\cite{Bauer:2017ris,Brivio:2017ije,Craig:2018kne}), where the loop factor in the anomaly coefficients in Eq.~\eqref{eq:ALPeft} is reabsorbed in the definition of the cut-off scale.

\section{Double production \& first order Electroweak Phase Transition}\label{sec:double}
When  $\sin^2\gamma$ goes below $10^{-3}$ single production becomes ineffective at CLIC. In this case, which is possibly related to an underlying $Z_2$ symmetry acting on the singlet, the only sizeable process can be the production of singlet pairs. This channel can give sizeable signals when the portal coupling $\lambda_{HS}$ in Eq.~\eqref{modello} is non negligible, as discussed in Section~\ref{sec:setup}. In this section we consider the following production processes
\be\label{doppia}
e^+ e^- \to \phi\phi\nu\bar\nu\ ,\qquad\qquad e^+ e^- \to \phi\phi e^{+}e^{-}\,,
\ee
which are the dominant ones at HELCs and can be relevant for singlet masses below 1 TeV at the Stage III of CLIC. The numerical value of the double production cross-section for $\sin\gamma=0$ is shown in the right panel of Figure~\ref{fig:rate-theory}. 

\begin{figure}[t]
\centering
\includegraphics[width=.49\textwidth]{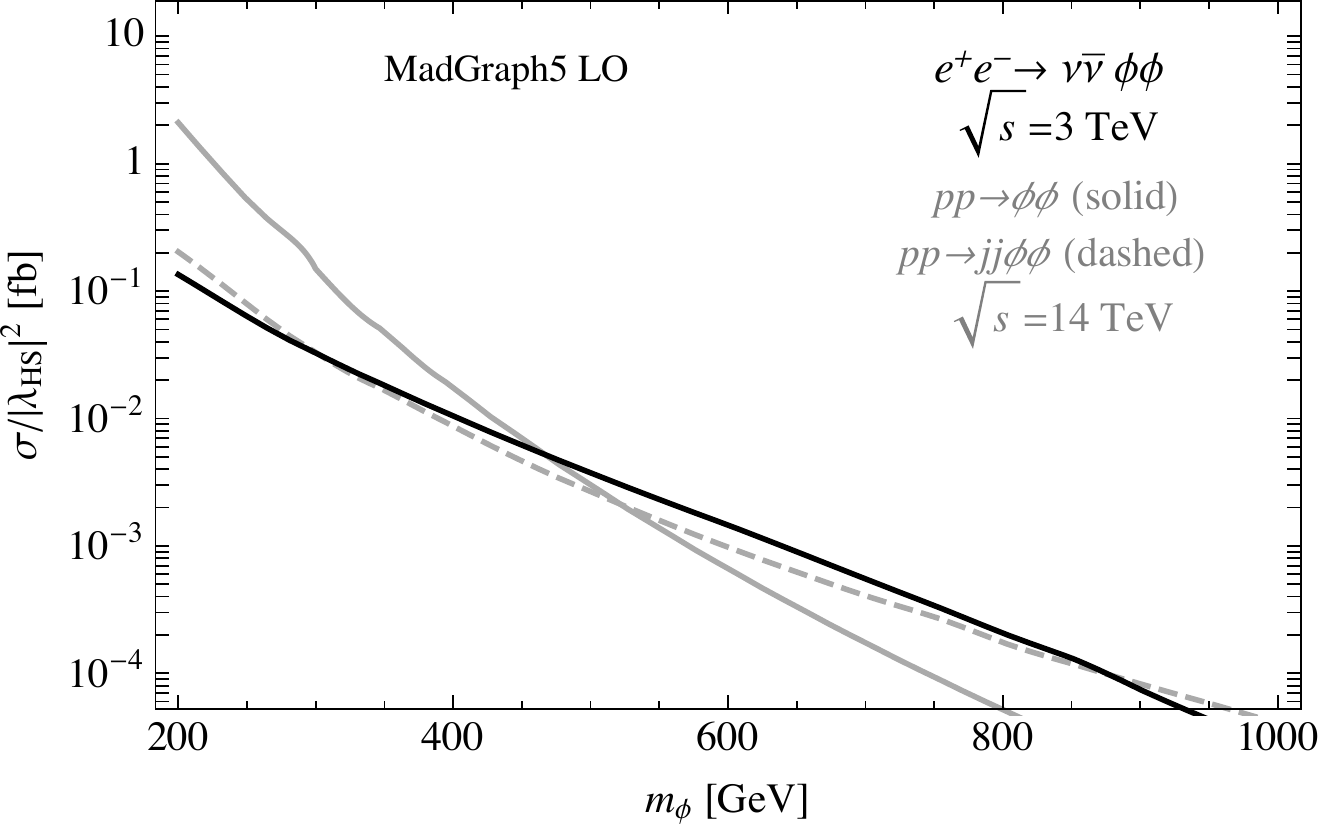} \hfill
\includegraphics[width=.49\textwidth]{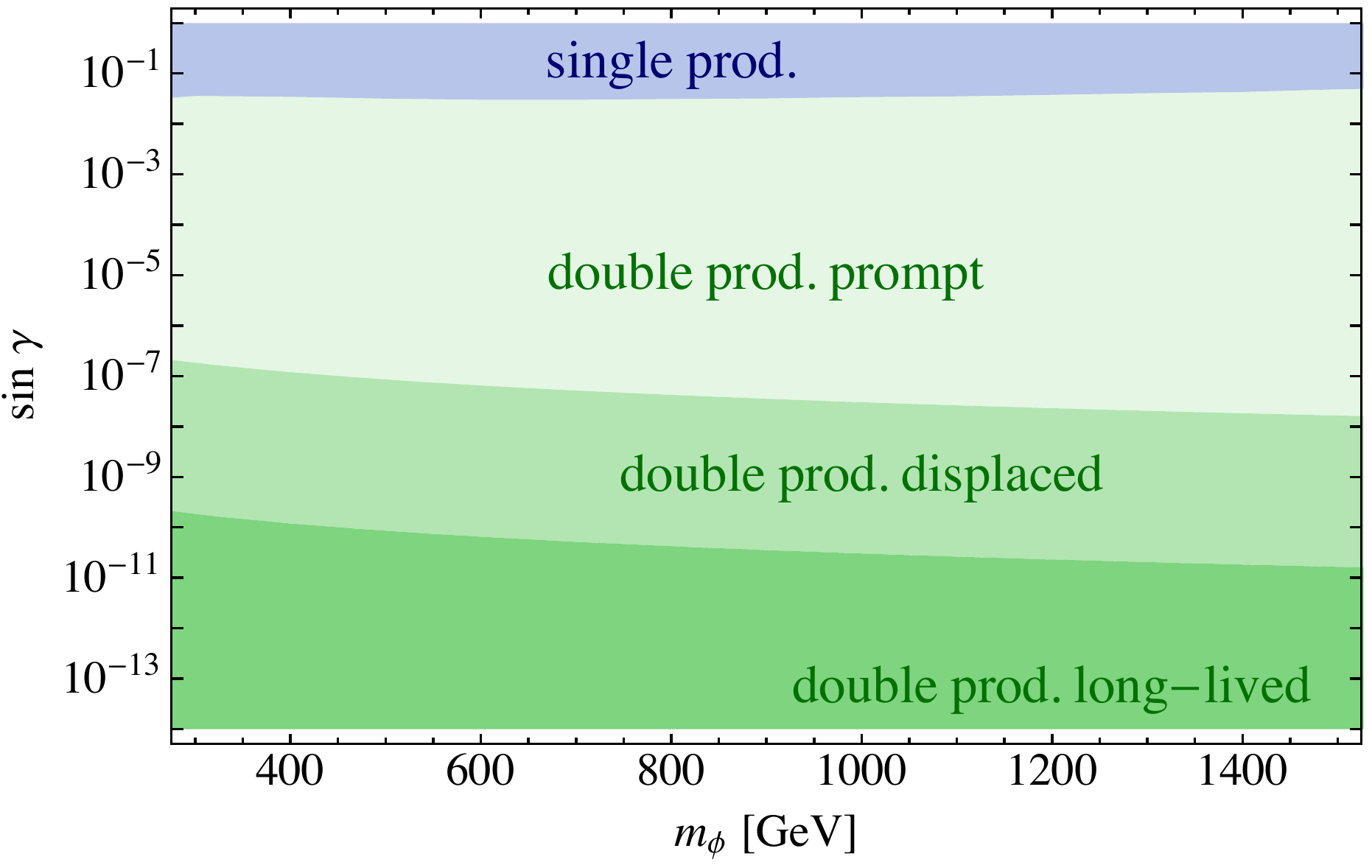} 
\caption{{\bf Left:} Pair production cross-section at the LHC and CLIC with normalisation $\lambda_{HS}=1$. {\bf Right:} A cartoon of the parameter space of the singlet model as a function of the mixing and the mass. The different green regions correspond to different singlet decay lengths: prompt $c\tau<0.1\text{ cm}$, displaced $0.1\text{ cm}<c\tau<100\text{ m}$ and long lived $c\tau>100\text{ m}$. We also shown in blue the reach in single production. \label{fig:doubleLHC}}
\end{figure}

In order to understand the possible sensitivities  of the channels in Eq.~\eqref{doppia} we first try to compare with the situation at the LHC, where related searches have been suggested. At LHC the VBF production can also benefit from sizeable couplings, however the gluon-fusion process is available and it dominates the total rate for light masses. This production channel has been exploited in a similar context in Ref.~\cite{Craig:2014lda,Chen:2017qcz}.\footnote{The total production cross-section can be computed by integrating over the gluon parton luminosities $dL_{gg}/d\tau$ as \cite{Plehn:1996wb}
\be
\sigma(pp\to \phi\phi)=\int_{4m_\phi^2/s}^1 d\tau\, \frac{dL_{gg}}{d\tau}\, \hat{\sigma}_{gg\to \phi\phi}(\hat{s}=\tau s)=\int_{4m_\phi^2/s}^1 d\tau\, \frac{dL_{gg}}{d\tau} \frac{\alpha_s^2|\lambda_{HS}|^2}{(2\pi)^3 512}\frac{\hat{s}|F(4m_t^2/\hat{s})|^2}{(\hat s - m_h^2)^2+\Gamma_h^2 m_h^2}\sqrt{1-\frac{4m_\phi^2}{\hat{s}}}\,,
\ee
where the loop function is $F(x)=x [1+(1-x)f(x) ]$, with $f(x)=-\frac{1}{4}\,(\log\frac{1+\sqrt{1-x}}{1-\sqrt{1-x}}-i\pi)^2$ for $\sqrt{\hat s}> 2m_t$. }
The comparison between the rate at the 14 TeV LHC and 3 TeV CLIC is plotted in the left panel of Figure~\ref{fig:doubleLHC}, where the rates are normalised to $\lambda_{HS}=1$. The typically smaller backgrounds at HELCs are likely to overcome the smallness of the rates for light masses. This suggests that lepton colliders can be an extremely useful tool to study pair production of scalar singlets that modify the Higgs potential.

Depending on the possible final states, different experimental strategies can be undertaken. A table with possible final states for a promptly decaying singlet and the relative branching fractions is given in Table~\ref{table:BRdouble}. Given the relatively small backgrounds for visible final states with multi jets and/or leptons, we could expect a large improvement when a combination of several channels is considered. For this reason in the left panel of Figure~\ref{fig:doubleLHC} we show a line of 100 signal events at 3 ab$^{-1}$ for $\lambda_{HS}=1$ in order to show the possible reach in mass if the singlets could be fully reconstructed. This corresponds to roughly 10 events in four vectors decaying hadronically.  

Broadly speaking, we can classify the final states in three main categories depending on the values of $\sin\gamma$, which controls the decay length of the scalar singlet. This gives rise to: $i)$ prompt decays; $ii)$ displaced decays; $iii)$ collider-stable singlets. The right plot in Figure~\ref{fig:doubleLHC} shows the various categories as a function of $\sin\gamma$. We now discuss the opportunities at HELC for these final states and the comparison with the analogous search strategies at the LHC.

\begin{table}
\renewcommand{\arraystretch}{1.15}
\centering
\begin{tabular}{r|c|c|c|c|c}
 $X(Y)$ & $4V(8j)$ &$2V(4j)2h(4b)$  &$2W(\ell\nu 2j)2h(4b)$&  $4h(8b)$ & $4W(3\ell 2j)$ \\ \hline
BR($\phi \to X \to Y$) & $0.12$ & $ 0.03 $ & $0.018$  & $7 \times 10^{-3}$ & $2.1 \times 10^{-3}$ 
 \end{tabular}
 \caption{A selection of branching ratios leading to interesting final states double singlet production, assuming ${\rm BR}(\phi \to WW) = 2\,{\rm BR}(\phi \to ZZ) = 2\,{\rm BR}(\phi \to hh) = 50\%$.\label{table:BRdouble}}
 \end{table}

\begin{itemize}
\item[$i)$] \emph{Prompt decays} are obtained in a large region of the singlet parameter space for which single production is subdominant. This corresponds to mixing angles between $10^{-2}$ and $10^{-6}$--$10^{-8}$. The latter numbers correspond to require a decay length of less than 1 mm for a singlet mass of 100 GeV and 1 TeV, respectively. In this portion of the parameter space multi-boson and multi-Higgs final states can give spectacular signals, with very low or even negligible backgrounds.

\item[$ii)$] \emph{Displaced decays} with displacement between 0.1 cm and 100 m are obtained for mixings down to $\sin\gamma\approx 10^{-8}-10^{-11}$.
This would lead to spectacular signals with multiple displaced tracks in the tracker (between 0.1 and 100 cm) or in the muon chamber (between 1 and few tens of meters).
This type of events has been shown to be basically background free at the current stage of the LHC (see e.g.\ Ref.~\cite{Aad:2015uaa,CMS:2014wda} and the sensitivity derived for displaced decays of scalar pairs in Ref.~\cite{Cui:2014twa}).
At HELC like CLIC these channels are going to be cleaner than at the LHC because the reduction in multi-jets backgrounds is going to diminish the expected rate of fake displaced vertices.  
\item[$iii)$]\emph{Invisible decays} correspond to mixings smaller than $10^{-11}$. This is the so-called ``nightmare scenario'' where the scalar singlet can only be pair produced and decay invisibly~\cite{Curtin:2012aa}. A comprehensive study on the possible reach at the LHC has been performed in Ref.~\cite{Craig:2014lda}. This shows that searches at the HL-LHC with $\sqrt{s}=14\text{ TeV}$ and 3 $\text{ab}^{-1}$ can probe $\lambda_{HS}\sim 6$ for $m_\phi\sim 200\text{ GeV}$ by combining double production channels for the singlet in VBF, gluon fusion and associated production with $t\bar{t}$. At CLIC the dominant background for an invisibly decaying singlet pair produced in $Z$-boson fusion is coming mostly from $\sigma(e^+e^{-}\to e^{+}\nu_e W^{-})=0.48\text{ pb}$ with the $W$ decaying leptonically.
By using our analytical estimate of the signal in Eq.~\eqref{double}, we can extrapolate at CLIC the results obtained in Ref.~\cite{Chacko:2013lna} for ILC at 1 TeV and 1000 $\text{fb}^{-1}$.
For a 100 GeV singlet we get $\frac{S}{\sqrt{B}} \simeq 2.3\,\lambda_{HS}^2$ at the stage III of CLIC, by rescaling the sensitivity in Ref.~\cite{Chacko:2013lna} with the square root of the luminosity times signal cross-section, thus assuming that the background cross-sections would scale in energy roughly as the signal ones.
This gives hope to cover this type of scenarios at CLIC. A more careful study for final states with missing transverse energy is left for future work.
\end{itemize}

In the next section we discuss how double production at HELCs can possibly probe a interesting region of the parameter space where a first order electroweak phase transition is induced by the singlet and the deviations in the Higgs couplings are below the best sensitivity of future lepton colliders.

\subsection{Electroweak phase transition}\label{sec:PT}

\begin{figure}[t]
\centering
\includegraphics[width=.65\textwidth]{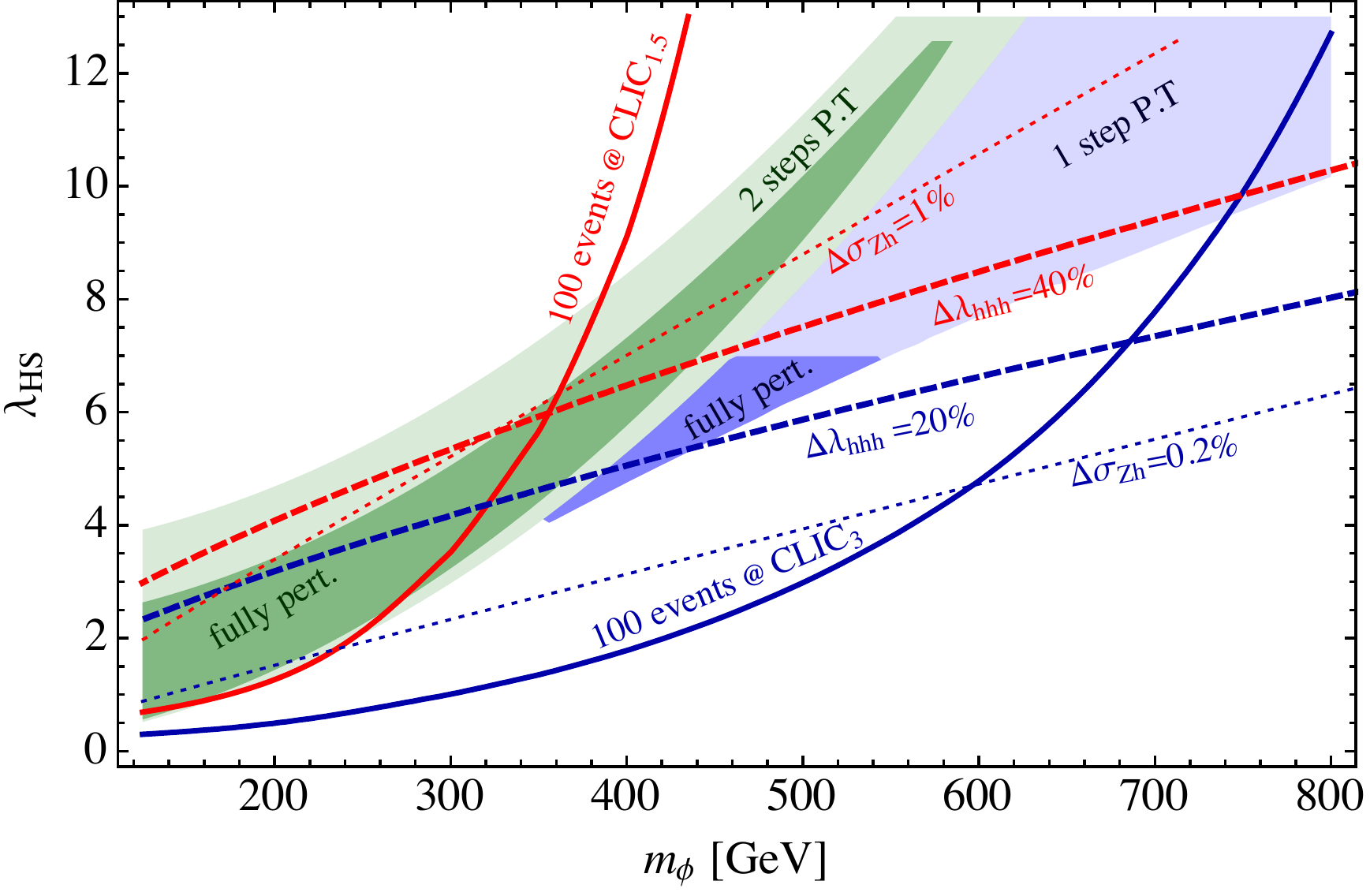} 
\caption{\label{fig:ewpt}Test of the double singlet production at CLIC. The region with a first order electroweak phase transition is enclosed in the shaded regions, as discussed in the text. Isolines of 100 events refer to $e^+ e^- \to \phi\phi\nu\bar\nu$ at CLIC 1.5 TeV (red) and CLIC 3 TeV (blue). Contour lines of deviations in the trilinear Higgs coupling (dashed, see Section~\ref{sec:setup}) and in $Zh$ production (dotted, see~\cite{McCullough:2013rea,Craig:2013xia}) correspond to possible CLIC sensitivities~\cite{Dawson:2013bba}.}
\end{figure}

It is well known that the presence of a singlet with a sizeable coupling to the Higgs field can induce a first order electroweak phase transition (FOEWPT) (see Ref.~\cite{Kotwal:2016tex} for a review and references). It is therefore natural to ask whether our analysis can constrain such scenarios thanks to the pair production of singlets.
Many detailed studies of the FOEWPT have been performed in the literature~\cite{Profumo:2007wc,Espinosa:2011ax,Chung:2012vg,peisi,Curtin:2012aa}, and numerical codes are available~\cite{Wainwright:2011kj}.
Here, we are willing to sacrifice numerical accuracy for analytic simplicity and we will adopt a as much as possible simplified description in order to characterise the main regions where a FOEWPT can occur.  

When the SM plasma has a sufficiently high temperature $T$, the electroweak symmetry is restored due to thermal correction to the Higgs mass.
Once the temperature drops as an effect of the expansion of the Universe, tunnelling to the true vacuum might happen. In what follows we describe the main features of the two regions where a FOEWPT can occur in our model, displayed in Figure~\ref{fig:ewpt}. To simplify our discussion, in this section we assume that the $Z_2$ symmetry forbidding the $a_S$ and $a_{HS}$ terms in the Lagrangian Eq.~\eqref{modello} is a good approximate symmetry, up to a small perturbation parametrised by the the mixing $\sin\gamma$, following Ref.~\cite{Curtin:2014jma}.   

\begin{itemize}
\item \emph{Two-step Phase Transition.} The so-called 2-step phase transition occurs when the singlet develops a VEV at finite temperature before the Higgs settles to its minimum. Later on (at smaller temperatures), this vacuum  tunnels via a strong first order PT to the vacuum with the present EW VEV, which is the global minimum for $T=0$.
The situation can be arranged with
\begin{equation}
m_S^2 = m_\phi^2-\lambda_{HS}v^2/2 < 0\,,
\end{equation}
and
$\lambda_S$ large enough, to guarantee that the global EWSB minimum has $\langle S\rangle=0$ while another local minimum appears at $h=0$ and $\langle S\rangle\neq0$. This condition gives a lower bound on $\lambda_S$: $\lambda_S\geq 2 m_S^4/m_h^2 v^2$.
When the above inequality is saturated the two minima are actually degenerate and the critical temperature approaches zero. All in all the region in the plane $(m_\phi,\lambda_{HS})$ where the two-step phase transition can occur is roughly given by
\be\label{2step}
m_\phi^2-\lambda_{HS} v^2/2 <0\,, \quad \mathrm{and}\, \quad \lambda_{S}v^2\frac{m_h^2}{2} \geq |m_\phi^2-\lambda_{HS} v^2/2|^2.
\ee
The requirement that the singlet develops a VEV before tunnelling at finite temperature to the vacuum with $\langle S\rangle = 0$ would actually impose a more stringent condition, the determination of which goes beyond the purpose of this paper. We choose to use the region defined by $m_S^2<0$, as it conservatively contains the region satisfying the more precise requirement.
The two different green regions in Figure~\ref{fig:ewpt} correspond to different perturbativity bounds on the couplings $\lambda_S$ and $\lambda_{HS}$: in the light green region we require $\beta_{\lambda_{HS}}/\lambda_{HS}<1$, while in the darker green region we also require $\beta_{\lambda_{S}}/\lambda_{S}<1$. $\beta_{\lambda_{HS}}$ and $\beta_{\lambda_{S}}$ are the beta functions of the corresponding couplings\footnote{The dominant contributions are $16\pi^2 \beta_{\lambda_{HS}}=\lambda_{HS}(12\lambda_H + 4\lambda_{HS} +6\lambda_S + 6 y_t^2)$, and $16\pi^2 \beta_{\lambda_{S}}=2\lambda_{HS}^2+18\lambda_S^2$.} (see Ref.~\cite{Goertz:2015nkp} for a discussion of the connection of this bound with the more standard requirement of not having Landau poles at the TeV scale). 

\item \emph{One-step Phase Transition.} As already discussed in Section~\ref{sec:setup}, in the $Z_2$-symmetric limit we can integrate out the heavy singlet  generating the dim 6 effective operators in Eq.~\eqref{matching}. In this scenario the quartic of the singlet does not play any role and can be set to zero for simplicity. A first order phase transition can occur in the Higgs EFT if the effective Higgs quartic is negative and the potential is stabilised by the dimension 6 operator $\vert H\vert^6$. In the $Z_2$-symmetric case, requiring the loop corrections to the Higgs quartic to be big enough to make it negative gives roughly the region displayed in Figure~\ref{fig:ewpt}. The precise shape of the region requires to include the full Coleman-Weinberg 1-loop contributions at zero temperature and the finite temperature corrections. We take this result from Ref.~\cite{Curtin:2014jma} which agrees sufficiently well with the computation that includes resummation of thermal loops \cite{Curtin:2016urg}. The perturbativity constraint requires $\beta_{\lambda_{HS}}/\lambda_{HS}<1$ as above. The singlet self-coupling is irrelevant in this scenario and can be set to zero. 
Notice that going beyond the $Z_2$ symmetric case, the first term of $c_6$ in Eq.~\eqref{matching} can be made large with a sizeable mixing angle. This scenario has been shown to induce a FOEWPT in Ref.~\cite{peisi1}. 
\end{itemize}

The main result of this section is Figure~\ref{fig:ewpt}, where we show the region where a FOEWPT occurs based on the discussion above.
Our approach has two main limitations.
First, we did not compute the temperature where bubble nucleation occurs, and this could further shrink the 2-step region as emphasised in Ref.~\cite{Kurup:2017dzf}. 
Second, we did not impose the condition for a fast decoupling of the sphaleron transitions inside the bubbles, $v_c/T_c \simeq 1$, which is a necessary requirement for EW baryogenesis. Such region, however, has been found to almost coincide with the blue shaded area in our Figure~\ref{fig:ewpt}, see e.g.\ Ref.~\cite{Curtin:2014jma}.
Therefore, we expect the region where EW baryogenesis may take place in this model to be fully contained in our shaded areas, at least for not too large values of the mixing angle $\sin\gamma$.

Figure~\ref{fig:ewpt} shows the relevance of pair production as a test of models with a FOEWPT.
For this purpose we plot isolines of 100 number of events at 1.5 TeV and 3 TeV CLIC. They may be enough to test this model under the reasonable assumption that many of the multi-Higgs and multi-bosons final states would face very small to zero backgrounds (e.g.\ $8b$), and that one could combine different channels as suggested by Table~\ref{table:BRdouble}. As shown in Figure~\ref{fig:doubleLHC}, lowering further the mixing angle would improve the reach of double production in the displaced region until the singlet would become long lived on collider scales. A dedicated analysis would be required to precisely assess the reach of CLIC for invisible final state. We also show the projected sensitivity of CLIC at 1.5 TeV and 3 TeV on triple Higgs coupling deviations (taken from Ref.~\cite{Dawson:2013bba}), and two benchmarks for deviations in $Zh$ associated production. These deviations are generically predicted in this setup as shown from Eq.~\eqref{eq:Higgscoupling}. 

In conclusion, pair singlet production at HELCs has the potential to test the entire parameter space allowing for FOEWPT and potentially EW baryogenesis. It constitutes a complementary probe to deviations in triple Higgs couplings and in $Zh$ associated production (see e.g.\ Ref.~\cite{Dawson:2013bba,DiVita:2017vrr}).

\section{Outlook and conclusions}\label{sec:conclusions}

A clean background environment and a high energy in the center of mass constitute a dream for any particle physicist.
Machines satisfying both properties are High Energy Lepton Colliders (HELCs), like CLIC~\cite{Linssen:2012hp} and more futuristic proposals of muon colliders~\cite{Antonelli:2015nla,Shiltsev:2018qbd,Alexahin:2013ojp,Delahaye:2013jla}.
In this paper we made progresses in building their physics case.

HELCs allow for powerful precision tests of the Higgs and electroweak sectors, as well as for direct production of new resonances beyond the reach of current experiments.
As discussed in the Introduction, the SM Higgs boson is dominantly produced in $WW$-fusion, which may allow a precision on the Higgs coupling at the per-mille level.
Building upon this observation, we studied direct $WW$-production of new scalar resonances which are singlet under the SM gauge group and couple to the SM through the Higgs sector. These particles are very well motivated from the theoretical view point and represent an important target for future collider machines.

The first message of our study is summarised in Figure~\ref{fig:comparison}. HELCs like CLIC extend the reach on the resonant production of a singlet decaying into di-bosons and/or di-Higgs well beyond the HL-LHC reach, i.e. down to couplings which correspond to Higgs coupling deviations at the per-mill level and up to 1--2 TeV masses. A similar conclusion holds for more ambitious proposals like future $\mu$ colliders (like LEMMA or MAP),  which stand as fantastic discovery machines to probe new physics coupled to the Higgs sector (see Figure~\ref{futuro} for a comparison with future hadron colliders). 
The consequences of the improved reach in single production are deep on the theoretical side, as we explored in Section~\ref{sec:models}. Singlet searches and Higgs coupling deviations might represent the only window to explore models of Neutral Naturalness like the Twin Higgs, where the coloured states can be beyond the LHC reach. In conventional models of Naturalness a new singlet might have other reason to be within the reach of CLIC, for example the SM Higgs mass in the NMSSM. The same searches can test heavy axion-like particles coupled to EW gauge bosons. These arise from the spontaneous breaking of approximate global symmetries and could be e.g.\ related with Dark Matter production and/or vector-like confinement. 

The second message of our study is shown in Figure~\ref{fig:ewpt}, where the reach in double production of singlet scalar particle coupled through the Higgs portal is compared against the allowed parameter space of electroweak baryogenesis. Depending on the decay length of the singlet different kinds of final states can be probed at CLIC. Without attempting to perform a detailed collider study for all of them, we point out that CLIC at $\sqrt{s}=3\,\TeV$ has the potential to explore models exhibiting a first order electro-weak phase transition, well beyond the reach of indirect constraints from Higgs coupling deviations. The same region can give correlated signals in future interferometers for gravitational waves such as LISA \cite{Caprini:2015zlo}. We believe that our results motivate a more detailed collider study of the various channels for singlet  pair production, in order to assess  robustly the reach of such mode. 

We hope that this study represents another little piece of motivation to push forward the quest to explore the high energy frontier.

\paragraph{Acknowledgements}~\\
{\small We thank Roberto Franceschini for the constant support and Ulrike Schnoor for helping us with the CLIC Delphes Card. We thank Roberto Franceschini and Andrea Wulzer for comments on the first version of this paper. We thank Jose Miguel No and Michael Spannowksy for correspondence on their forthcoming work \cite{josemi}. DB is supported by the grant ``FLAVOR'' from the INFN. FS is partly supported by a PIER Seed Project funding (Project ID PIF-2017-72).
AT is partially supported by the grant ``STRONG" from the INFN. FS and AT are grateful to the MITP for hospitality and partial support during the completion of this work.}


\appendix

\section{Details on the $hh(4b)$ analysis}\label{app:4b}

Here we provide some further details on the analysis performed in Section~\ref{sec:Stohh} in the $\phi\to hh\to 4b$ channel.

\begin{figure}
\centering%
\includegraphics[width=0.49\textwidth]{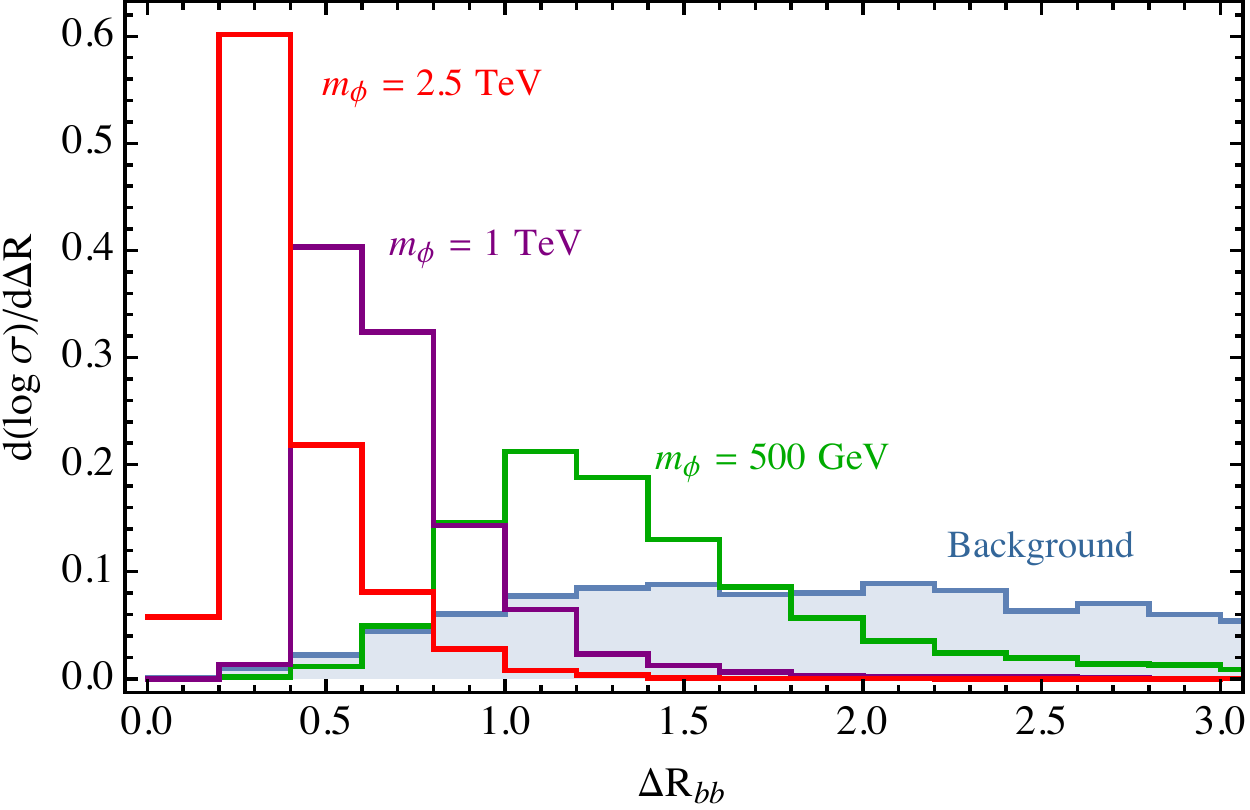}\hfill%
\includegraphics[width=0.49\textwidth]{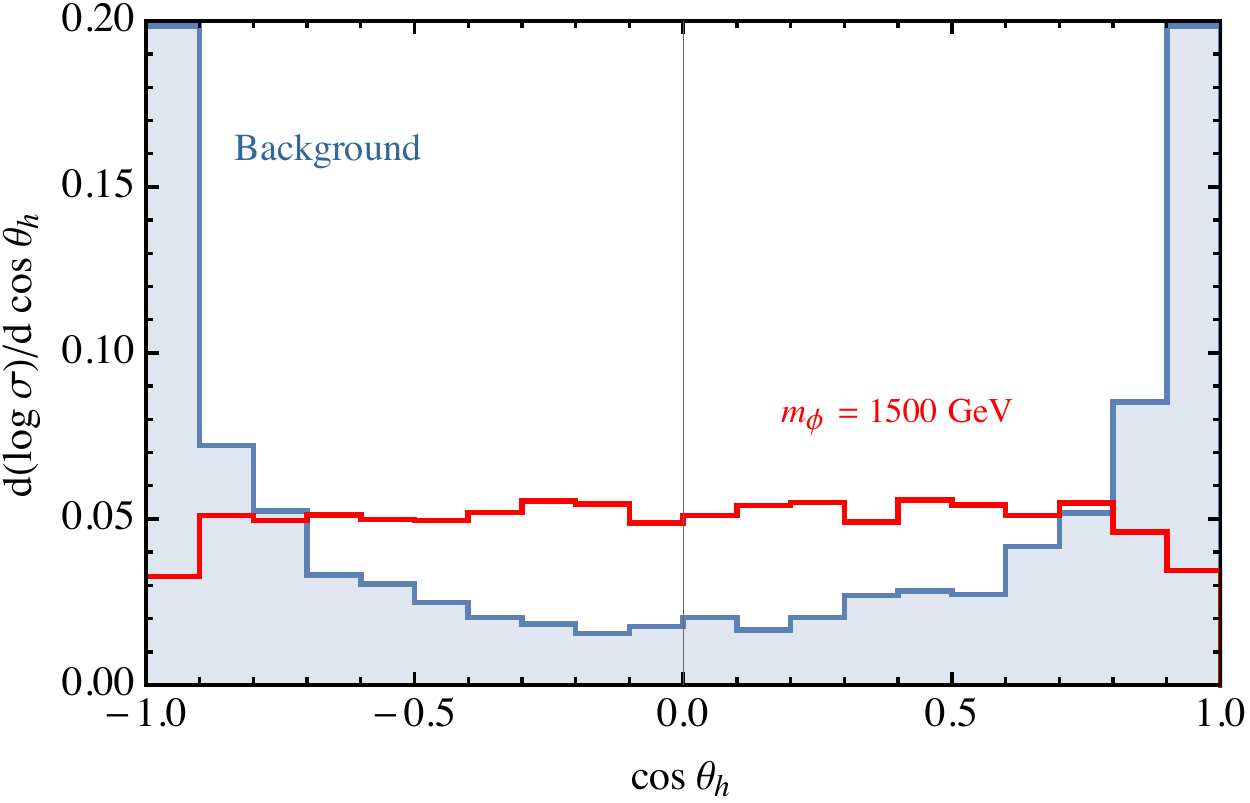}
\caption{{\bf Left:} Distribution in $\Delta R_{bb}$ of the background and the signal (for $m_\phi = 500$ GeV, 1 TeV, and 2.5 TeV) at the 3 TeV CLIC. {\bf Right:} Distribution in $\cos\theta_h$ for both signal ($m_\phi = 1$ TeV) and background at 3 TeV.\label{app1}}
\end{figure}

The plot on the left-hand side of Figure~\ref{app1} shows the distribution in $\Delta R_{bb}$ of the two $b$-jets that reconstruct the Higgs bosons. It can be seen that for higher masses of the resonance the two Higgs bosons are more boosted, resulting in lower values of $\Delta R_{bb}$. The CLIC detectors are expected to be able to resolve separations down to at least $\Delta R_{jj}\approx 0.1$, so we can exploit the full power of the exclusive jet reconstruction algorithm (which in our case considers $\Delta R$ low enough to identify $N=4$ jets) up to the kinematical limit at 3 TeV. Notice that the algorithm is able to reconstruct 4 jets in virtually all the signal events that we simulated. The distribution for the background, instead, has jets that are typically separated by larger values of $\Delta R_{bb}$.

The plot on the right-hand side of Figure~\ref{app1} shows instead the distribution in the cosine of the polar angle $\theta_h$ of the Higgs bosons, for signal and background. We can see how the background peaks in the forward region, $\cos\theta_h > 0.8$, while the distribution of the signal is flat.

\section{Details on the sensitivities at higher energies}\label{app:lepton}

\begin{figure}[t!]
\centering%
\includegraphics[width=0.99\textwidth]{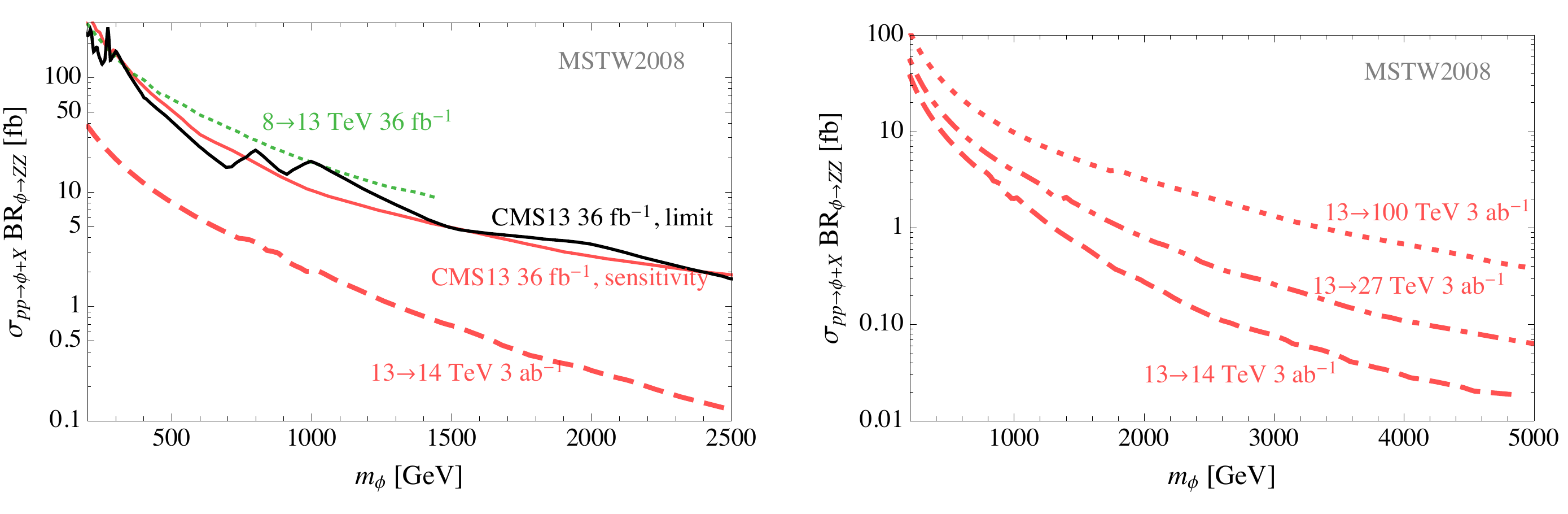}
\caption{\label{fig:XS_hadrons} Sensitivities on the signal cross-sections of a new resonance decaying into $ZZ$, at the high-luminosity (dashed) and high-energy (dot-dashed) LHC, and at the FCC-hh (dotted). They are derived from the present LHC sensitivity at 13 TeV with 36 fb$^{-1}$ of luminosity~\cite{Sirunyan:2018qlb} (continuous red), see text for more details. For comparison, we also show the 13 TeV limit corresponding to the sensitivity of Ref.~\cite{Sirunyan:2018qlb} (continuos black), and the sensitivity that one would obtain at 13 TeV with 36 fb$^{-1}$, rescaling the 8 TeV searches~\cite{Khachatryan:2015cwa}. The latter serves as a rough validation of our procedure.}
\end{figure}

\paragraph{\bf Future hadron colliders.}
We show in Figure~\ref{fig:XS_hadrons} the reach in cross-section for a singlet that decays to pairs of vector bosons at the HL-LHC, the HE-LHC and at FCC-hh.
These are the sensitivities that we used to draw the lines in Figures~\ref{fig:comparison} and \ref{futuro}.
As described in Section~\ref{sec:hadron}, we determine them by rescaling the expected sensitivity of the existing 13~TeV search~\cite{Sirunyan:2018qlb} to higher energies and luminosities using quark parton luminosities, with a procedure analogous to the one presented in Ref.~\cite{Buttazzo:2015bka}.
At the time Ref.~\cite{Buttazzo:2015bka} was published only 8~TeV LHC results were available, so the sensitivities presented there were obtained rescaling the 8~TeV searches.
The results of Figure~\ref{fig:XS_hadrons} are thus to be considered as an update of Ref.~\cite{Buttazzo:2015bka}. As a rough validation of the method, we find that the 13 TeV sensitivity determined rescaling the 8~TeV results~\cite{Khachatryan:2015cwa} with the same procedure is in reasonable agreement with the one of the actual experimental search~\cite{Sirunyan:2018qlb} (see the green dashed and continuous red lines in Figure~\ref{fig:XS_hadrons}).

\begin{figure}[t!]
\centering%
\includegraphics[width=0.49\textwidth]{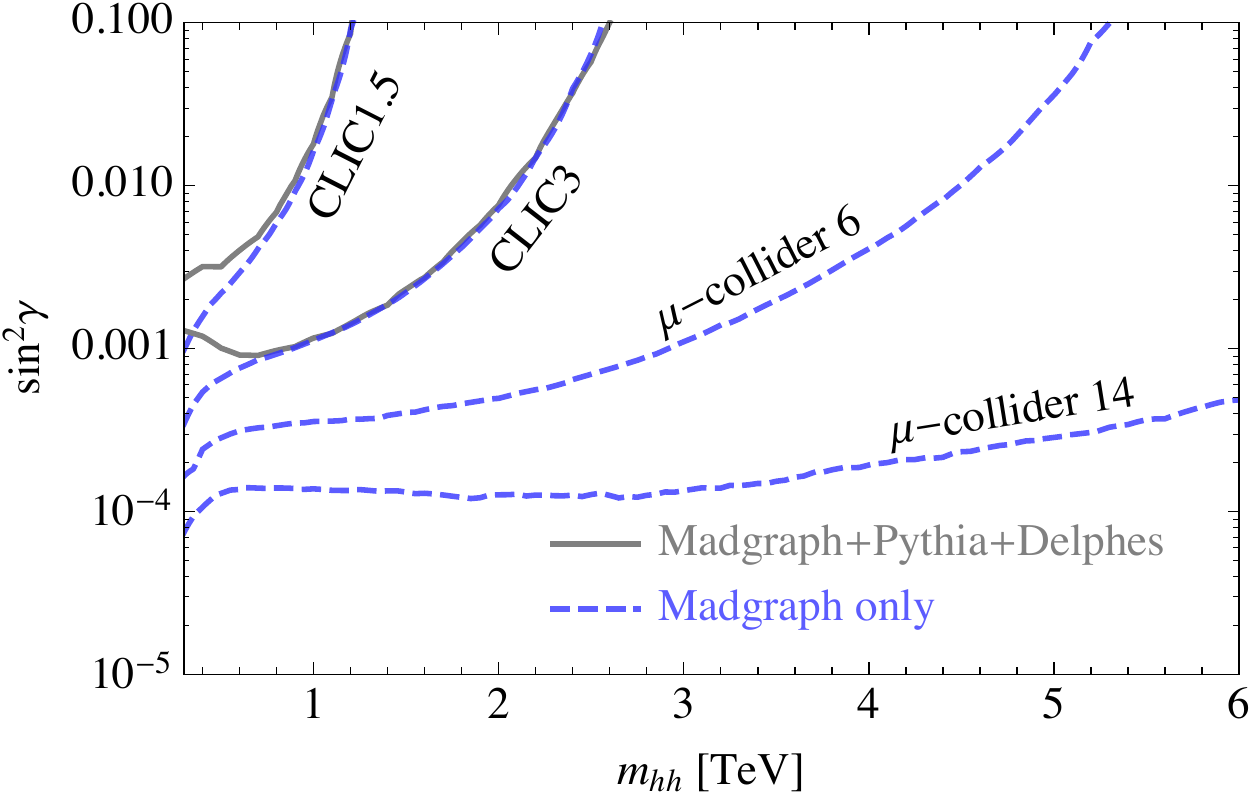}\hfill%
\includegraphics[width=0.49\textwidth]{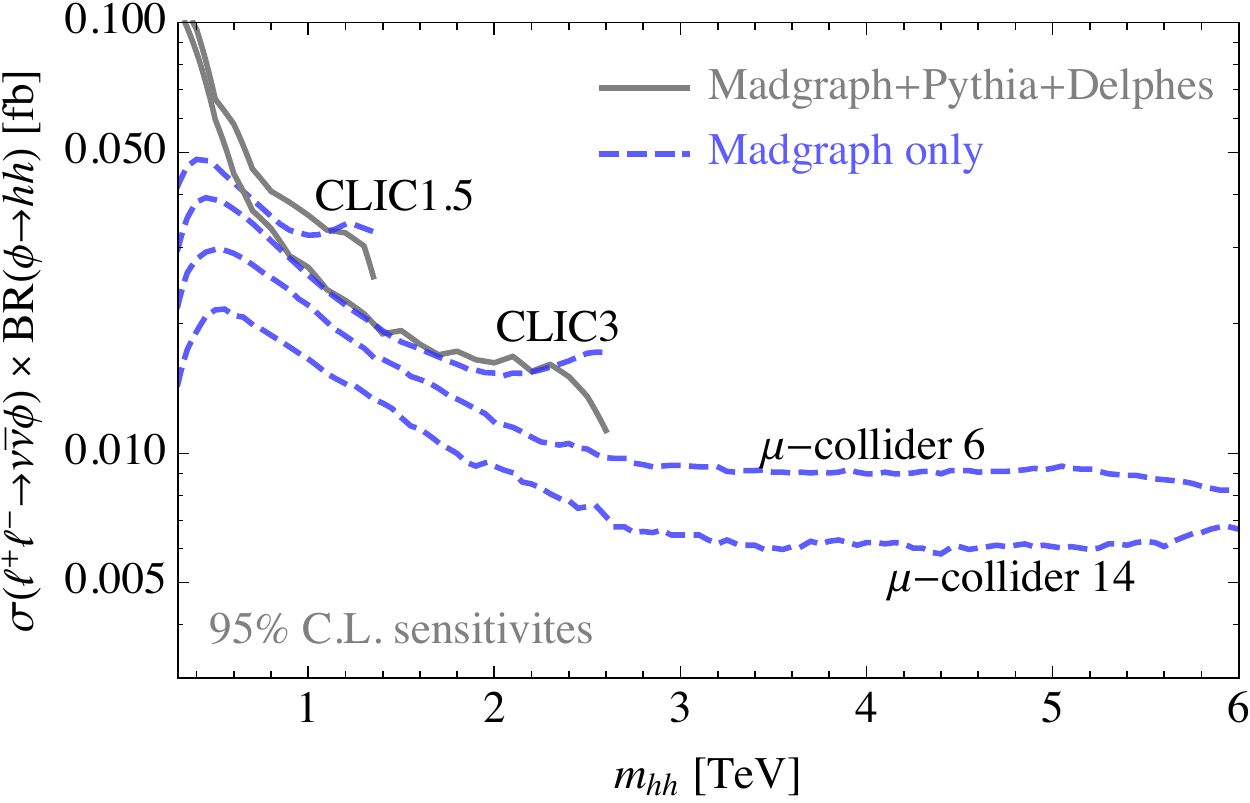}
\caption{\label{fig:MGvsDelphes} Comparison between the determination of sensitivities using {\sc Madgraph} only (both at CLIC and future muon colliders) and the full results for CLIC, as a validation of our simplified analysis. {\bf Left:} Higgs-singlet mixing angle, assuming ${\rm BR}(\phi\to hh) = 25\%$. {\bf Right:} signal cross-section of a generic resonance produced in $e^+e^-$ and decaying to $hh$.}
\end{figure}

\paragraph{\bf Future lepton colliders.}
In Figure~\ref{fig:MGvsDelphes} we plot the sensitivities on the signal cross-sections of a singlet that decays to pairs of Higgs bosons, at CLIC Stage II (1.5~TeV, 1.5~ab$^{-1}$) and Stage III (3~TeV, 3~ab$^{-1}$), and at a $\mu$-collider 6 (6~TeV, 6~ab$^{-1}$) and $\mu$-collider 14 (14~TeV, 14~ab$^{-1}$). They are determined at the {\sc Madgraph} level, from the simulation of the background $\ell^+\ell^- \to \nu \bar{\nu} hh$ at parton level, as explained in Section~\ref{sec:muon_colliders}.
The comparison of the sensitivities determined in this way (dashed blue lines), with those determined from a proper study including detector simulation in Section~\ref{sec:Stohh} (continuous gray lines), gives an explicit visualisation of the agreement between the two.

\begin{figure}[t!]
\centering%
\includegraphics[width=0.48\textwidth]{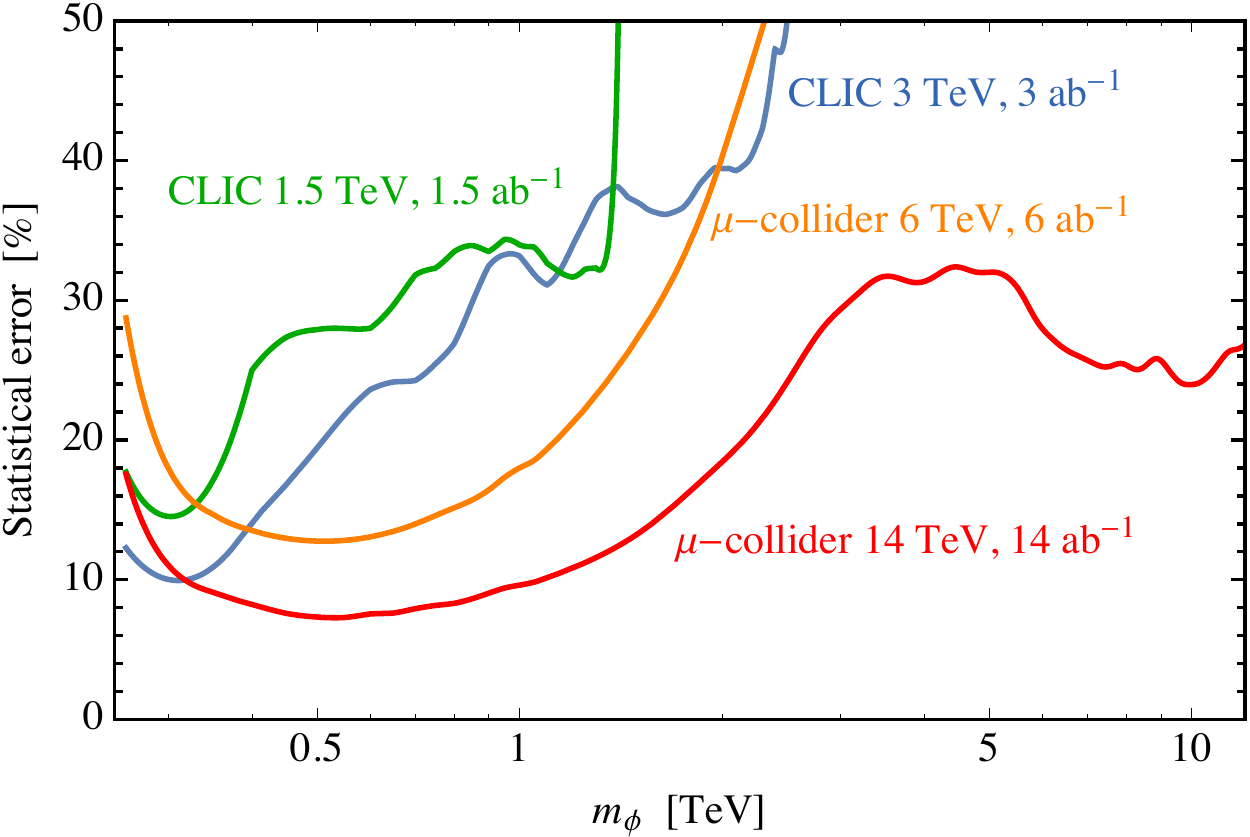}\hfill%
\includegraphics[width=0.5\textwidth]{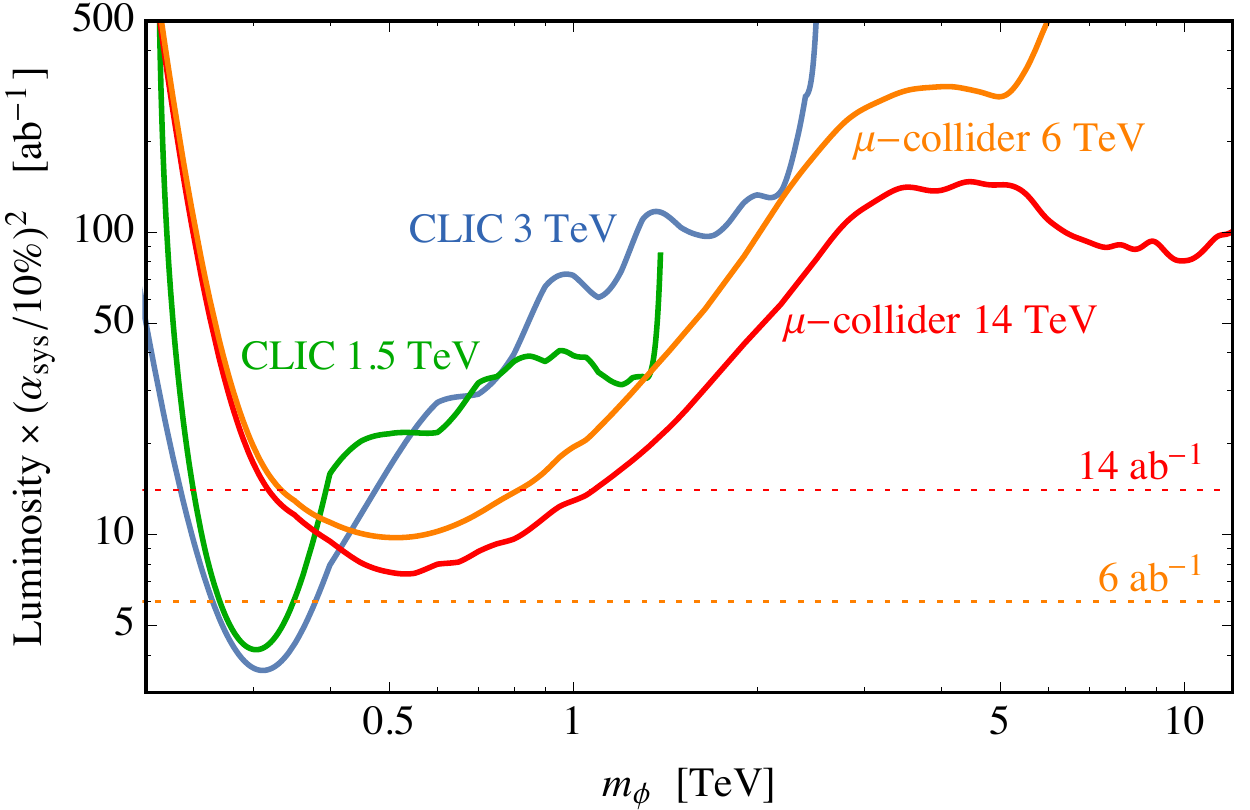}
\caption{\label{fig:stat-syst} {\bf Left:} Statistical uncertainty on the background expected in resonant $hh$ searches at CLIC and muon colliders. {\bf Right:} luminosity at which the impact of statistical uncertainties on our sensitivities becomes equal to the one due to systematics (above the continuous lines a systematic uncertainty of 10\% is more important than the statistical one).}
\end{figure}

In order to better assess the robustness of our analysis, we also compare the relative importance of statistical and systematic uncertainties in setting our limits and sensitivities.
On the left-hand side of Figure~\ref{fig:stat-syst} we show the statistical error of the sensitivities at CLIC and at muon colliders, for the specific values of the luminosities that we have used. On the right-hand side of the same figure, we show the value of the luminosity needed in order for the systematic uncertainties to become important. For definiteness, we use a conservative benchmark value of $\alpha_{\rm sys} = 10\%$, recalling that the resulting luminosity scales as $\alpha_{\rm sys}^{-2}$.
These results show that resonant $hh$ searches at CLIC will always be statistically dominated, and thus our results are independent of the precise value of $\alpha_{\rm sys}$ used in the analysis.
Similarly, for the chosen benchmark luminosities, our estimated reaches at muon colliders are largely dominated by statistical errors for masse above a TeV, while they are expected to become sensitive to systematic errors of $\sim 15\%$ (6~TeV) and $\sim 7\%$ (14~TeV) for low masses.


\pagestyle{plain}
\bibliographystyle{jhep}
\small
\bibliography{BRST}

\end{document}